\documentclass[aps,twocolumn,prd,preprintnumbers,superscriptaddress,nofootinbib,bibnotes,floatfix,longbibliography]{revtex4-1}
\pdfoutput=1
\usepackage[colorlinks=true,citecolor=blue,linkcolor=blue,breaklinks=true]{hyperref}
\usepackage{amsmath,amssymb,amsfonts,stackengine}
\usepackage{epsfig}  
\usepackage{graphicx}
\usepackage{url}
\usepackage{bm}
\usepackage{multirow}
\usepackage{placeins}
\usepackage[dvipsnames]{xcolor}
\usepackage{ulem}

\usepackage{float}
\usepackage{comment}

\allowdisplaybreaks
\bibpunct{[}{]}{,}{n}{}{,}

\newcommand{\w}{\omega}
\newcommand{\W}{\Omega}
\newcommand{\R}{\varrho}
\newcommand{\tx}{\text}
\newcommand{\Gf}{G_{\tx{F}}}

\begin{document}
\title{Fast neutrino flavor conversion, ejecta properties, and nucleosynthesis in newly-formed hypermassive remnants of neutron-star mergers}

\author{Manu George}
\email{kuttan.mgc@gmail.com}
\affiliation{Institute of Physics, Academia Sinica, Taipei, 11529, Taiwan}

\author{Meng-Ru Wu}
\email{mwu@gate.sinica.edu.tw}
\affiliation{Institute of Physics, Academia Sinica, Taipei, 11529, Taiwan}
\affiliation{Institute of Astronomy and Astrophysics, Academia Sinica, Taipei, 10617, Taiwan}
\affiliation{Physics Division, National Center for Theoretical Sciences, 30013 Hsinchu, Taiwan}

\author{Irene Tamborra}
\email{tamborra@nbi.ku.dk}
\affiliation{Niels Bohr International Academy and DARK, Niels Bohr Institute, University of Copenhagen, Blegdamsvej 17, 2100, Copenhagen, Denmark}

\author{Ricard Ardevol-Pulpillo}
\email{ricardar@mpa-garching.mpg.de}
\affiliation{Max-Planck-Institut f\"ur Astrophysik, Postfach 1317, 85741 Garching, Germany}

\author{Hans-Thomas Janka}
\email{thj@mpa-garching.mpg.de}
\affiliation{Max-Planck-Institut f\"ur Astrophysik, Postfach 1317, 85741 Garching, Germany}

\date{\today}
\begin{abstract}
Neutrinos emitted in the coalescence of two neutron stars affect the dynamics of the outflow ejecta and the nucleosynthesis of heavy elements.
In this work, we analyze the neutrino emission properties and the conditions leading to the growth of flavor instabilities in merger remnants consisting of a hypermassive neutron star and an accretion disk during the first 10~ms after  the merger. The analyses are based on hydrodynamical simulations that include a modeling of neutrino emission and absorption effects via the ``improved leakage-equilibration-absorption scheme'' (ILEAS). 
We also examine the nucleosynthesis of the heavy elements via the rapid neutron-capture process ($r$-process) inside the material ejected during this phase. 
The dominant emission of $\bar\nu_e$ over $\nu_e$ from the merger remnant leads to favorable conditions for the occurrence of fast pairwise flavor conversions of neutrinos, independent of the chosen equation of state or the mass ratio of the binary. 
The nucleosynthesis outcome is very robust, ranging from the first to the third $r$-process peaks. 
In particular, more than $10^{-5}$~$M_\odot$ of strontium are produced in these early ejecta that may account for the GW170817 kilonova observation.
We find that the amount of ejecta containing free neutrons after the $r$-process freeze-out, which may power early-time UV emission, is reduced by roughly a factor of 10 when compared to simulations that do not include weak interactions. Finally, the potential flavor equipartition between all neutrino flavors is mainly found to affect the nucleosynthesis outcome in the polar ejecta within $\lesssim 30^\circ$, by changing the amount of the produced iron-peak and first-peak nuclei, but it does not alter the lanthanide mass fraction therein.
\end{abstract}
\maketitle

\section{Introduction}
Compact binary systems consisting of two neutron stars (NS) or a NS and a black hole (BH) can lose their angular momentum through continuous emission of gravitational waves (GW), eventually leading to the merging of the compact objects.
Such merger events have long been considered to be the sites producing short gamma-ray bursts (sGRB) and synthesizing heavy elements via the rapid neutron-capture process ($r$-process)~\cite{1974ApJ...192L.145L,Eichler:1989ve,Cowan:2019pkx}, 
which powers electromagnetic transients in optical and infrared wavelengths, the so-called kilonovae~\cite{Li:1998bw,Kulkarni:2005jw,Metzger:2010sy,Metzger:2016pju}.

The first detected GW emission from a binary neutron star merger event by the LIGO and Virgo Collaborations (GW170817) 
together with multi-wavelength electromagnetic observations have confirmed theoretical predictions~\cite{TheLIGOScientific:2017qsa,Monitor:2017mdv,GBM:2017lvd}.
Future observations, like the one of GW170817, will be able to offer further opportunities to precisely determine the population of binary NS systems 
and the yet-uncertain rich physics involved in binary NS mergers, including the nuclear equation of state (EoS) and the properties of neutron-rich nuclei.
In order to achieve these goals, solid theoretical modeling of mergers is needed without any doubt.

The interaction of neutrinos with matter and their flavor conversions in the binary NS merger environment are among the most uncertain theoretical aspects that can affect the observables.
A copious amount of neutrinos and antineutrinos can be produced by the merger as matter is heated up to  several tens of MeV due to the collision of two NSs.
Neutrinos play an important role in determining the cooling of the merger remnant, 
changing the composition of the ejecta, and altering the $r$-process outcome and the kilonova emission properties. 

The early-time blue color of the GW170817 kilonova and the recently inferred amount of strontium production~\cite{Watson:2019xjv} both suggest that 
the merger ejecta contain some less neutron-rich material with electron fraction per nucleon $Y_e\gtrsim 0.3$.
On the other hand, numerical simulations which include weak interactions of neutrinos with nucleons
all suggest that neutrino emission and absorption have the
effect to reduce the neutron-richness of the outflow launched at different post-merger phases, preferably in the direction perpendicular to the merger plane~\cite{Wanajo:2014wha,Perego:2014fma,Fernandez:2013tya,Sekiguchi:2015dma,Radice:2016dwd,Foucart:2016rxm,Fujibayashi:2017puw,Miller:2019dpt,Vincent:2019kor}.
In particular, recent work suggests that neutrino absorption can be responsible for increasing $Y_e$ even for the early-time dynamical ejecta in the polar direction~\cite{Wanajo:2006ec,Sekiguchi:2015dma,Radice:2016dwd,Goriely:2016cxi,Ardevol-Pulpillo:2018btx,Miller:2019dpt}. 
Meanwhile, the $\nu_e \bar\nu_e$ pair annihilation to $e^+e^-$ pairs above the accretion disk, although it might not be the
dominant driver, can also contribute to launch the sGRB jet~\cite{Eichler:1989ve,Woosley:1993wj,Ruffert:1998qg,2011MNRAS.410.2302Z,Just:2015dba}

Neutrinos additionally undergo flavor conversions above the merger remnant, altering the neutrino absorption rates on nucleons as well as their pair-annhilation rates~\cite{Malkus:2012ts,Malkus:2014iqa,Wu:2015fga,Zhu:2016mwa,Frensel:2016fge,Wu:2017qpc,Wu:2017drk}, thus possibly affecting the interpretation of the observed signals.
In particular, Ref.~\cite{Wu:2017qpc} showed that favorable conditions for the so-called ``fast flavor conversion''~\cite{Sawyer:2005jk,Sawyer:2015dsa,Dasgupta:2016dbv,Izaguirre:2016gsx,Capozzi:2017gqd,Abbar:2017pkh,Yi:2019hrp,Capozzi:2019lso,Chakraborty:2019wxe} exist nearly everywhere
above the merger remnant because of the disk geometry and the protonization of the merger remnant in its effort to reach a new beta-equilibrium state for the high temperatures produced in the merger process. Fast pairwise conversions can  give rise to rapid flavor oscillations of neutrinos within a length scale of  
$\sim(\Gf| n_{\nu_e}-n_{\bar\nu_e}|)^{-1}$ $\approx\mathcal{O}(1)$~cm. 
Subsequently, Ref.~\cite{Wu:2017drk} adopted time-dependent neutrino emission characteristics from simulations of merger remnants consisting of a central BH with an accretion disk and also found favorable conditions for fast flavor conversion. By assuming full flavor equilibration between all neutrino flavors, it was shown that the nucleosynthesis outcome in the neutrino-driven outflow from the BH--disk remnant can be largely altered~\cite{Wu:2017drk}.

In this work, we focus on the newly formed hypermassive remnants of NS mergers. In fact, in the case of a hypermassive NS, which might transiently exist in many binary NS mergers, the surrounding torus-like equatorial bulge (``disk'') is exposed to strong neutrino irradiation from the massive core of the merger remnant. This is different from the BH-torus configuration, where there is no such intense central neutrino source. Moreover, the torus itself produces heavy-lepton neutrinos only with very low luminosities, whereas the hypermassive NS radiates high luminosities also of the heavy-lepton neutrinos. Both aspects make it necessary to investigate the question of fast flavor conversions for the case where the hypermassive NS with a surrounding disk/torus still exists.
To this purpose, we  focus on the first $\simeq 10$~ms after the merger of two NSs, during which a central hypermassive NS surrounded by an accretion disk forms.
By relying on recent hydrodynamical simulations with two different EoS and two different mass ratios for the binary, developed in Refs.~\cite{Ardevol:thesis,Ardevol-Pulpillo:2018btx} and available at~\cite{Ardevol2019}, we explore  the neutrino emission properties, the conditions for the occurrence of fast  flavor instabilities, and the effects of neutrino absorption and flavor conversions on the neutron-richness of the ejecta as well as the nucleosynthesis outcome.

The paper is organized as follows.
In Sec.~\ref{sec:model}, we introduce the merger remnant models and the neutrino emission properties. 
In Sec.~\ref{sec:flav_instability}, we outline the framework used for the linear stability analysis  and present our numerical results on the occurrence of fast flavor conversions. 
We analyze the ejecta properties, the neutrino absorption effects on the  evolution of $Y_e$, and the impact of flavor equipartition in Sec.~\ref{sec:nucleo}.
We summarize our findings and discuss their implications in Sec.~\ref{sec:conclusion}.
We adopt natural units with $\hbar=c=1$ for all equations throughout the paper.

\section{Neutrino emission from binary neutron star mergers}\label{sec:model}

\subsection{Models of binary neutron star mergers}
\label{sec:simulation}
We consider models of binary NS mergers simulated by a three-dimensional relativistic smoothed particle hydrodynamics code that adopts conformal flatness conditions~\cite{Bauswein:2013yna}.
The code has recently been coupled to an approximate neutrino transport scheme called ``improved-leakage-equilibration-absorption scheme (ILEAS),'' 
which is implemented on a three-dimensional (3D) Cartesian coordinate grid ($x, y ,z$) with the axis perpendicular to the merging plane chosen as $z-$axis. ILEAS serves as an efficient transport method for multidimensional simulations; when compared to two-moment neutrino transport results for protoneutron stars and post-merger tori,  
it captures well  neutrino energy losses from the densest regions of the system as well as neutrino absorption in the free-streaming regime~\cite{Ardevol-Pulpillo:2018btx,Ardevol:thesis}.

Our fiducial models discussed in this section are from Ref.~\cite{Ardevol-Pulpillo:2018btx} and simulate the mergers of two non-rotating NSs with mass 1.35 $M_\odot$ each, with the EoS of DD2~\cite{Typel:2009sy,Hempel:2011mk} and SFHo~\cite{Steiner:2012rk}.
In addition, we consider in later sections cases of unequal mass binaries consisting of two NSs with 1.25~$M_\odot$ and 1.45~$M_\odot$ each, based on the same numerical scheme of Refs.~\cite{Ardevol-Pulpillo:2018btx,Ardevol:thesis}.
We note here that we have taken all simulation data averaged over the azimuthal angle.
This is a fairly good assumption as the merger remnant quickly reaches an approximately axi-symmetric state after $\sim 2-3$~ms.

\begin{figure}
    \centering
    \includegraphics[width = \linewidth]{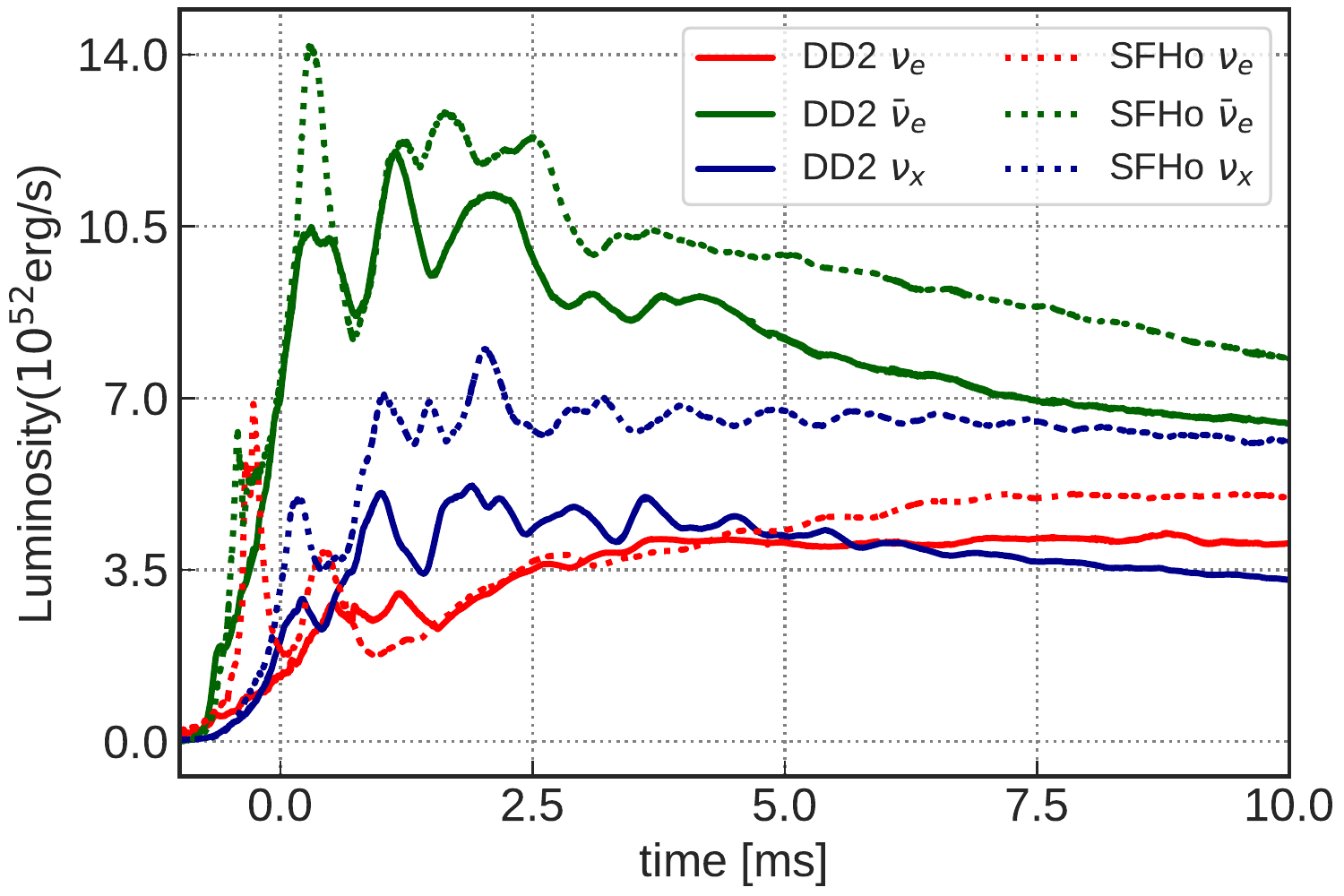}
    \includegraphics[width = \linewidth]{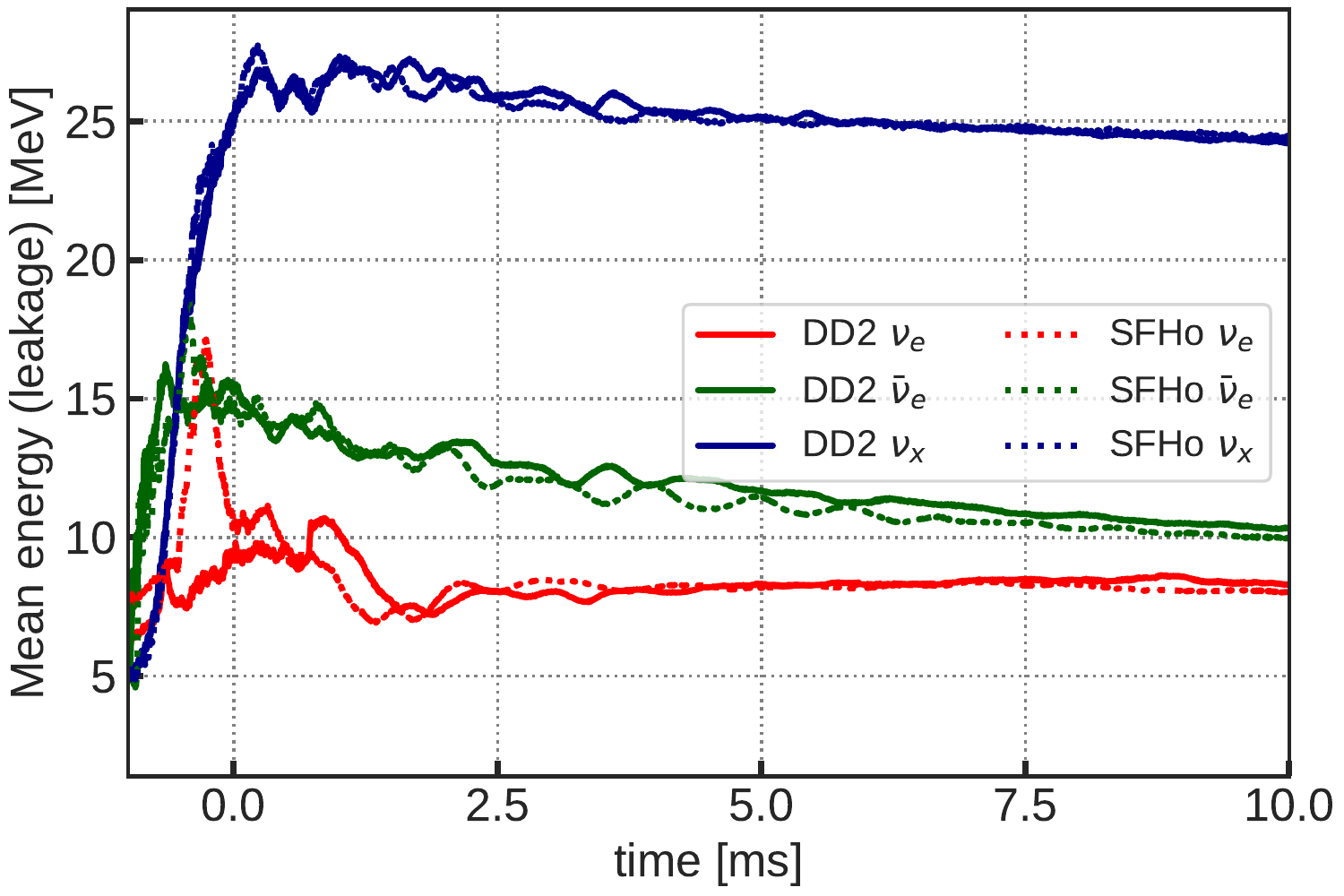}
    \caption{Time evolution of the neutrino energy luminosity and average energy for different species, $\nu_e$ (in red), $\bar\nu_e$ (in green) and $\nu_x$ (in blue) of the $1.35+1.35$~$M_\odot$ NS-NS merger simulation, with DD2 (solid lines) and SFHo (dashed lines) EoS. 
    }
    \label{fig:lumegy}
\end{figure}

\begin{figure*}[ht]
    \includegraphics[width = \linewidth]{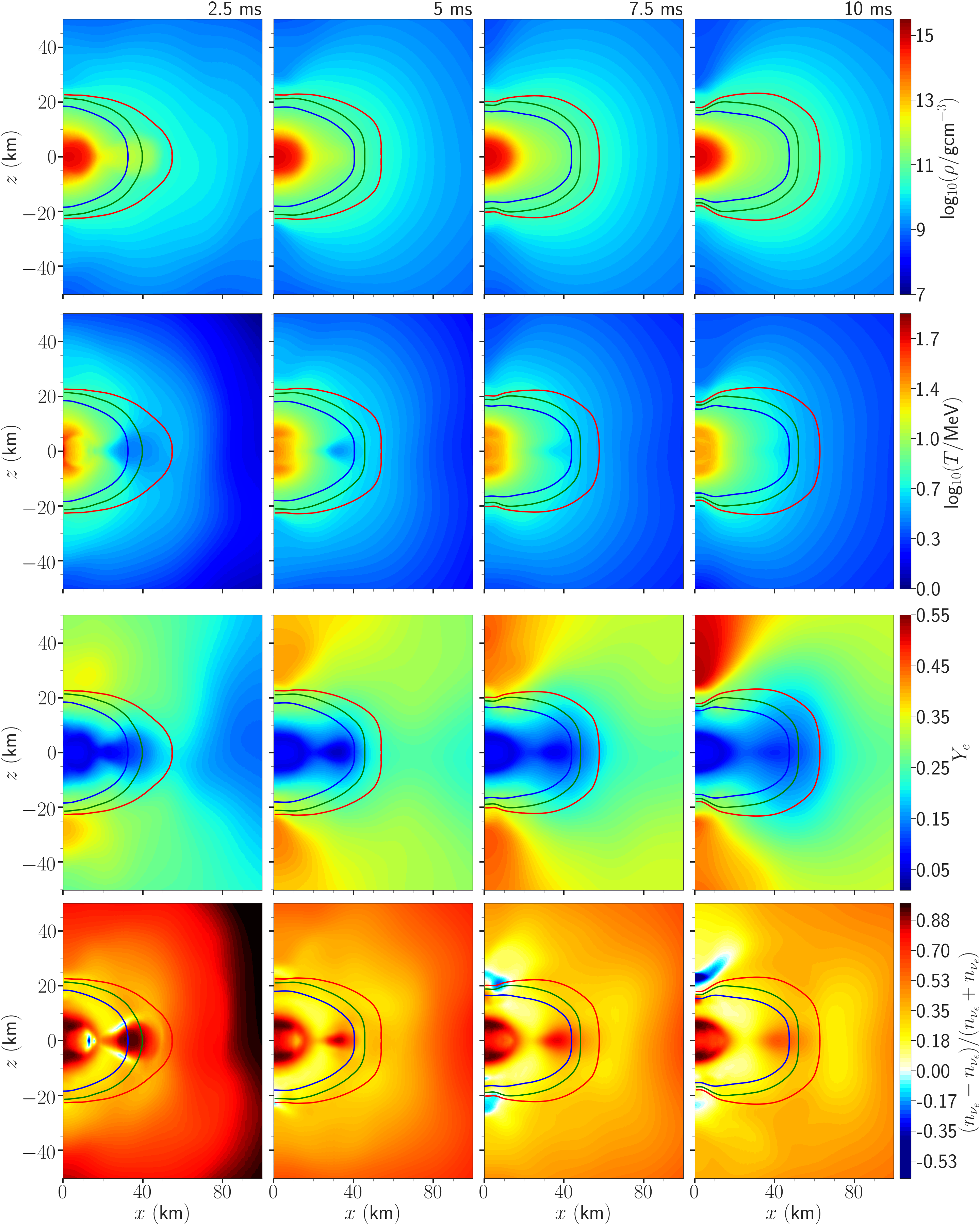}
    \label{fig:DD2sima}
    \caption{Hydrodynamic and neutrino properties from the $1.35+1.35$~$M_\odot$ NS-NS merger remnant simulation with DD2 EoS in the ($x,z$) plane for the time snapshot taken at  2.5, 5, 7.5, and 10 ms   after the coalescence (from left to right columns). 
    The first, second and third rows show the baryon mass density $\rho$, temperature $T$,  the electron number fraction $Y_e$. The fourth row displays the ratio
    $(n_{\bar\nu_e}-n_{\nu_e})/(n_{\bar\nu_e} + n_{\nu_e})$ . 
    A value of this ratio above (below) 0 implies larger $n_{\bar\nu_e}$ ($n_{\nu_e}$) locally.  
    Also shown with solid lines are the contours of the $\nu_e$ (red),  $\bar\nu_e$ (green), and $\nu_x$ (blue) emission surfaces.}
    \label{fig:simsnapDD2}
\end{figure*}
\begin{figure*}[ht]
    \includegraphics[width = \linewidth]{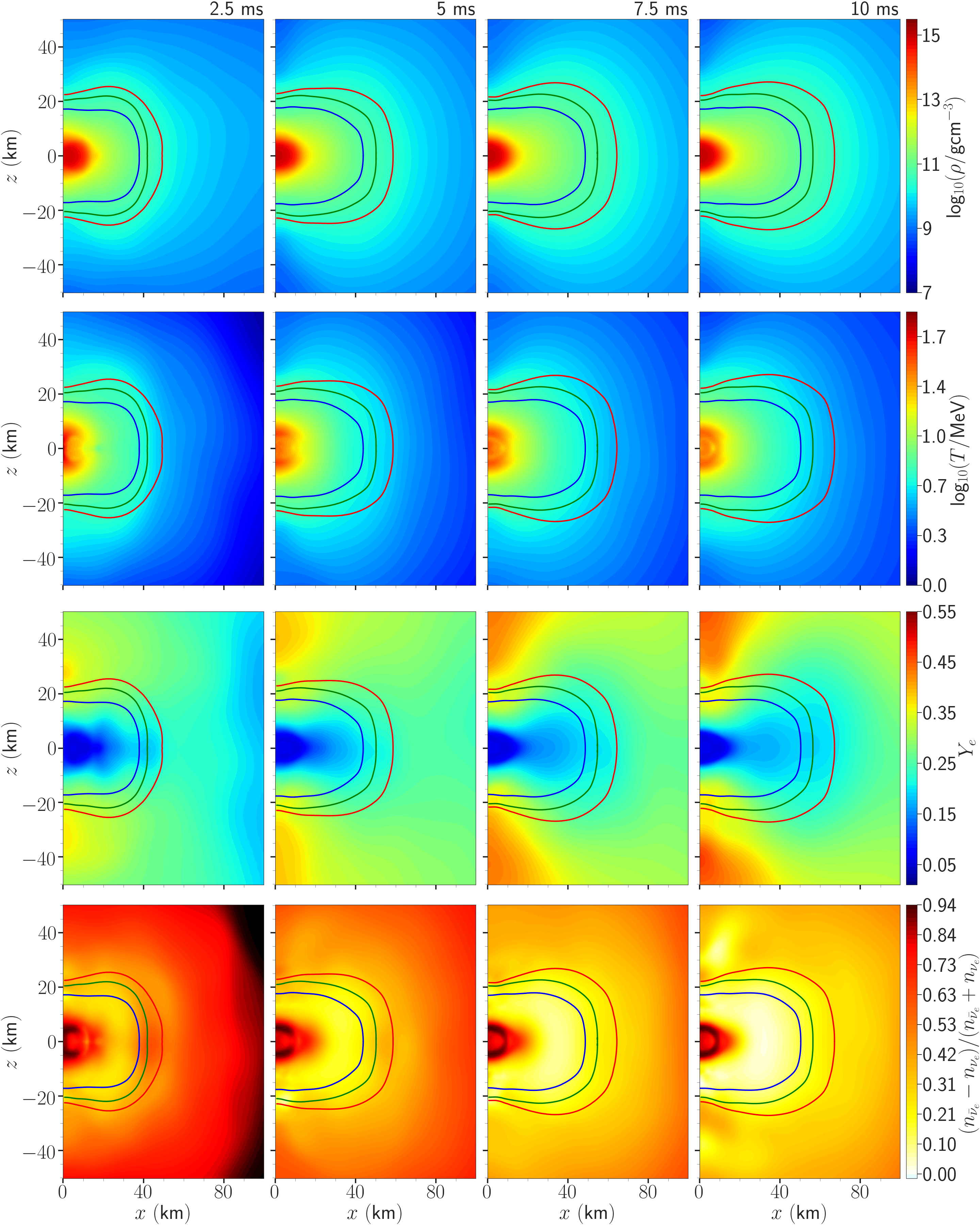}
    \caption{Same as Fig.~\ref{fig:simsnapDD2}, but for the simulation with SFHo EoS.
    }
    \label{fig:simsnapSFHo}
\end{figure*}

Figure~\ref{fig:lumegy} shows the evolution of the energy luminosity of different neutrino species and their average energy during the first 10~ms after the merger of two NSs for our symmetric mergers~\footnote{More precisely, the time is measured with respect to the first minimum of the lapse function (see Ref.~\cite{Ardevol-Pulpillo:2018btx}).}.
During this time, the central object is a hypermassive NS supported by differential rotation.
The energy luminosity of $\bar\nu_e$ ($L_{\bar\nu_e}\simeq 10^{53}$~erg~s$^{-1}$)
is a factor of 2 or 3 larger than the one of $\nu_e$ throughout the whole 10~ms, leading to continuous protonization of the remnant. 
The energy luminosities of the heavy-lepton neutrino flavors, denoted by $\nu_x$ (with $\nu_x$ being representative of one species of heavy-lepton neutrinos),
are $\sim 5\times 10^{52}$~erg~s$^{-1}$ per species.
The averaged energies estimated by the leakage approximation show a clear hierarchy of 
$\langle E_{\nu_e}\rangle < \langle E_{\bar\nu_e}\rangle < \langle E_{\nu_x}\rangle$,
reflecting the ordering of the temperature at decoupling; 
neutrinos that decouple in the innermost region of the remnant have higher average energy (see below and Fig.~\ref{fig:simsnapDD2}).
When comparing the energy luminosity evolution of the models with DD2 and SFHo EoS, 
one sees that the latter produces higher 
luminosities in all flavors.
This is related to the fact that the SFHo EoS is softer than the DD2 EoS, 
which results in a NS with smaller radius.
Consequently, a more violent collision during the merging leads to higher temperatures, and correspondingly higher neutrino emission rates, when the SFHo EoS was adopted.
This gives rise to a faster protonization of the remnant.

Comparing Fig.~\ref{fig:lumegy} to Fig.~2 of Ref.~\cite{Wu:2017drk}, where the latter displays the neutrino emission properties for a BH remnant, 
one can see that the neutrino emission properties of the electron-flavor neutrinos are comparable in the two scenarios. 
However, the non-electron-flavor neutrinos are more abundant in the models investigated in this work and have average energies higher than the electron-flavor neutrinos. 
In addition, the neutrino energy luminosities in the case with the BH accretion disk quickly decrease after $\simeq 20$~ms (see Fig.~2 of Ref.~\cite{Wu:2017drk}). 
Although we only focus on the first 10~ms after the coalescence, the neutrino energy luminosity reaches a plateau in the models where a hypermassive NS forms.

The first three rows in Fig.~\ref{fig:simsnapDD2} and Fig.~\ref{fig:simsnapSFHo} show the baryon density, temperature, and electron fraction profiles  
in the half $x-z$ plane for $x>0$ at 2.5, 5, 7.5, and 10~ms post merger
for the models with DD2 and SFHO EoS, respectively. 
Also shown are the locations of the $\nu_e$, $\bar\nu_e$, and $\nu_x$ emission surfaces.
The emission surface for a given neutrino
species $\nu_\alpha$ ($\nu_e$, $\bar\nu_e$, or $\nu_x$) is defined by a surface where the energy-averaged optical depth is $\tau_{\nu_\alpha}=2/3$
(see Sec.~\ref{sec:nu_surf} for details).
One can see from the density profiles that the remnant consists of a rotating deformed hypermassive NS surrounded by an accretion disk. 
The central object retains its initial low $Y_e\lesssim 0.1$ (see the third rows) and it is surrounded by a dense and neutron-rich disk,  
which is opaque to neutrinos for an extended radius out to $\sim 40-60$~km.
Both $\nu_e$ and $\bar\nu_e$ decouple at locations where the matter density is approximately between $10^{11} - 10^{12}$~g~cm$^{-3}$ and the temperature is $\sim 5$~MeV.
The $\bar\nu_e$ emission surfaces generally sit inside the ones of $\nu_e$ during 
this period, independently of the adopted EoS. 

The size of the neutrino emission surfaces slightly expands as the remnant evolves,
due to the settling of the post-merger object and the redistribution of matter with high angular momentum toward the equatorial plane, where the disk formation process proceeds. 
Moreover, as the remnant keeps protonizing,  $Y_e$ inside the neutrino surfaces
gradually increases with time.
The thick disk of the remnant in the model with the softer SFHo EoS protonizes faster as discussed above,
and thus has higher $Y_e$ inside the disk when compared to the profiles with DD2 EoS. 
Note that in the polar region close to the $z$-axis, a high $Y_e\gtrsim 0.4$ funnel forms
at later times, as both the differences between the $\nu_e$ and $\bar\nu_e$ luminosities and
average energies become smaller.

In the bottom panels of both Fig.~\ref{fig:simsnapDD2} and Fig.~\ref{fig:simsnapSFHo}, we show the ratio of the difference of the number densities of $\bar\nu_e$ and $\nu_e$  to the
sum of the $\bar\nu_e$ and $\nu_e$ densities, 
i.e., $(n_{\bar\nu_e}-n_{\nu_e})/(n_{\bar\nu_e} + n_{\bar\nu_e})$. 
Once again, as the remnant is protonizing and emitting more
$\bar\nu_e$ than $\nu_e$,
nearly any location  above the $\nu_e$ surface in both models has this ratio larger than zero during the entire first 10~ms.
The only exception is represented by the small patches in the polar region at 7.5 and 10~ms for the model with DD2 EoS.
These patches are a consequence of the neutronization that takes place locally around the poles of the high-density core of the merger remnant in the DD2 case. Because the stiff EoS prevents the merger core from further contraction, the polar regions cool quickly, and the density just inside the neutrinospheres increases by gravitational settling, forcing the neutrinospheres to move inward to smaller radii. This explains the more pronounced polar trough of the neutrinospheres in the DD2 model compared to the SFHo merger. Striving for a new beta-equilibrium state, now at lower temperature, the plasma begins to neutronize again, radiating more electron neutrinos than antineutrinos in both polar directions. This leads to the excess of $\nu_e$ relative to $\bar\nu_e$ outside the neutrinospheres, visible as the two blue patches in the two bottom right panels of Fig.~\ref{fig:simsnapDD2}. Correspondingly, $\nu_e$ captures dominate $\bar\nu_e$ captures in this region and the polar outflow becomes more and more proton-rich (red regions of $Y_e > 0.5$ around the $z$-axis in the right panels of the third row of Fig.~\ref{fig:simsnapDD2}). The same trends are visible in the SFHo simulation (Fig.~\ref{fig:simsnapSFHo}), though less extreme and less rapidly evolving than in the DD2 run.

\subsection{Neutrino number densities on their emission surfaces}
\label{sec:nu_surf}

The inner regions of the merger remnant are dense enough to trap neutrinos. 
This allows us to define a neutrino emission surface for each species $\nu_\alpha$ 
above which $\nu_\alpha$ can approximately free-stream. 
As we will use the properties of $\nu_e$ and $\bar\nu_e$ on their respective surfaces to
construct their angular distributions outside the $\nu_e$ surface in Sec.~\ref{sec:flav_instability},
we discuss below the time evolution and the dependence on the adopted EoS of the neutrino densities on their emission surfaces.

For any point $\bm x$ inside the simulation domain, the optical depth along a specific path 
$\gamma$ to another point $\bm y$ that a neutrino with an energy $E$ traverses is given by 
\begin{equation}
    \tau_{\nu_\alpha}(E, \bm x, \gamma) = \int_{\bm x}^{\bm y} \lambda^{-1}_{\nu_\alpha}(E, \bm x') ds,
    \label{eq:odepth}
\end{equation}
where $\bm x^\prime$ is a point, $ds$ is the differential segment along $\gamma$, 
and $\lambda_{\nu_\alpha}(E, \bm x'(s))$ is the corresponding mean-free-path at $\bm x^\prime$.
For each species $\alpha$, we determine the position of the neutrino decoupling surface in the same way as in 
Ref.~\cite{Ardevol-Pulpillo:2018btx}, i.e.~for every point $\bm x$, the minimum of the spectral-averaged $\langle\tau_{\nu_\alpha}(\bm x,i)\rangle$ is computed along the six different directions ($i \in(\pm x, \pm y, \pm z)$) through the edge of the simulation domain. 
The neutrino decoupling surface is then defined by the location corresponding to $\langle\tau_{\nu_\alpha}(\bm x,i)\rangle = 2/3$ in the six directions. 
The emission surfaces computed in this way for $\nu_e$, $\bar\nu_e$, and $\nu_x$ are the ones shown in Figs.~\ref{fig:simsnapDD2} and \ref{fig:simsnapSFHo}.

\begin{figure*}[ht]
    \includegraphics[width = \linewidth ]{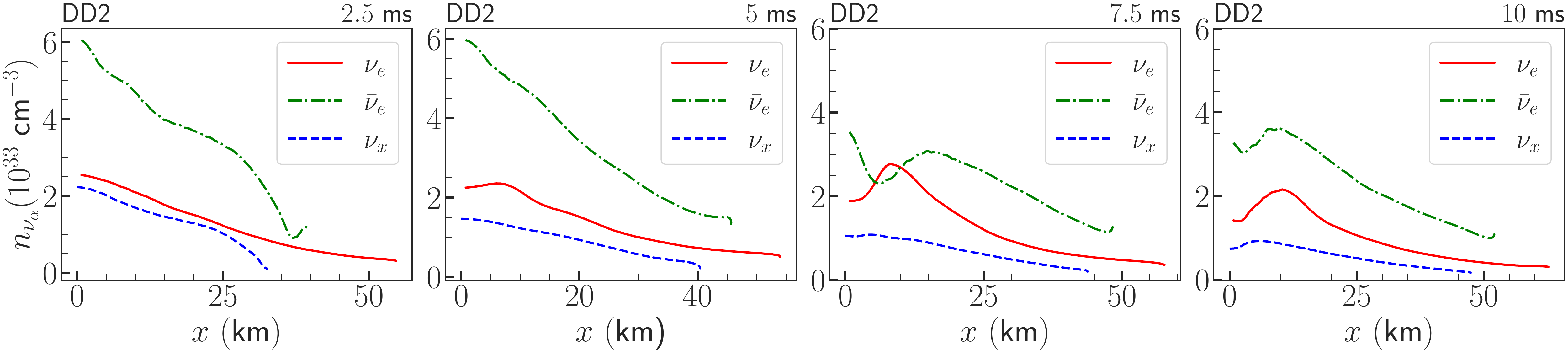}\\
    \includegraphics[width = \linewidth ]{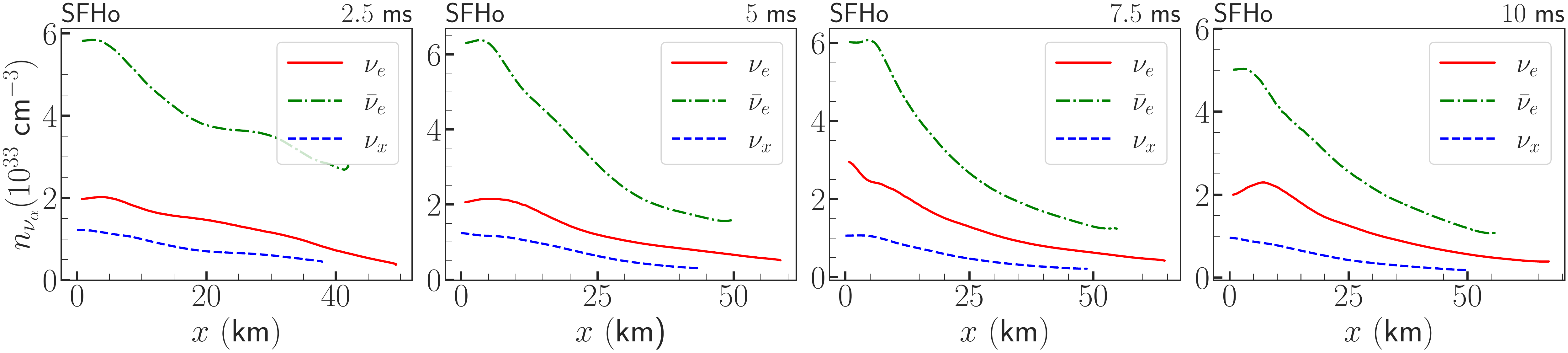}
    \caption{Number density of $\nu_e$ (red solid curves), $\bar\nu_e$ (green dash-dotted curves), and $\nu_x$ (blue dash lines) 
    on their respective emission surfaces in the northern hemisphere for the $1.35+1.35$~$M_\odot$ merger models with DD2 EoS (top panels) and SFHo EoS (bottom panels) 
    at 2.5, 5, 7.5 and 10~ms post merger.
    Note that the $\nu_x$ number density has been rescaled, see text and Appendix~\ref{app:postproc} for details.
    }
    \label{fig:neudensities}
\end{figure*}

Figure~\ref{fig:neudensities} shows the number densities of $\nu_e$, $\bar\nu_e$, and $\nu_x$
at their respective emission surfaces as functions of $x$ for different snapshots. 
Note that the $\nu_e$ and $\bar\nu_e$ number densities are computed by using  Fermi-Dirac distribution functions with temperature and  chemical potential extracted at each location on the neutrino surfaces, consistently with  Ref.~\cite{Ardevol-Pulpillo:2018btx}. 
For $\nu_x$, we rescale the number density according to the procedure described in Appendix~\ref{app:postproc} to account for the trapping effect between the energy-surfaces and the emission surfaces, which is known to considerably reduce the $\nu_x$ luminosity when compared to the analogous values directly estimated through the local emission (see, e.g., \cite{Janka:1995cu}).
The top panels show the neutrino density corresponding to the DD2 EoS, while the bottom panels display the same quantities for the SFHo EoS. 
As discussed in Sec.~\ref{sec:simulation}, the softer SFHo EoS leads to a generally higher
temperature in the merger remnant; as a consequence, the $\bar\nu_e$ luminosity (Fig.~\ref{fig:lumegy}) and number density on the emission surface are higher than 
 the one obtained in the model with DD2 EoS.
Both $n_{\nu_e}$ and $n_{\bar\nu_e}$ are larger in a region close to the pole
where the temperature is higher (see Figs.~\ref{fig:simsnapDD2} and \ref{fig:simsnapSFHo}).
Of relevance to the occurrence of fast flavor conversions is the fact that  the
 $\bar\nu_e$ emission  is  a factor of 2--3 larger than the one of $\nu_e$ across the whole 
emission surface; it remains quite stable throughout the simulated evolution until 10 ms after the plunge.
The only exception is the snapshot at 7.5~ms for the model with DD2 EoS,
which shows a dip in the $\bar\nu_e$ number density around $x \simeq 10$~km. 
This is because the enhanced deleptonization above the poles of the high-density core (described in Sec.~\ref{sec:simulation}) leads to a short, transient excess of $\nu_e$ over $\bar\nu_e$ near the neutrinospheres in the northern hemisphere (see Fig.~\ref{fig:simsnapDD2}). The effect is somewhat pathological and unusual, also because it is considerably less strongly developed in the southern hemisphere.
Notably, comparing Fig.~\ref{fig:neudensities} with the bottom panel of Fig.~1 of Ref.~\cite{Wu:2017drk}, one can see that the neutrino-antineutrino asymmetry is more pronounced in the present models than in the BH remnant case.

\section{flavor Instability}
\label{sec:flav_instability}

In Sec.~\ref{sec:dispersion} we first briefly introduce the theoretical formalism, viz., the dispersion relation (DR) approach~\cite{Izaguirre:2016gsx}, 
widely used in the literature to investigate the occurrence of neutrino flavor conversions. 
In Sec.~\ref{sec:ELN} we look for the conditions leading to flavor instabilities using the simulation data introduced in Sec.~\ref{sec:model}.
We then apply the DR formalism to investigate the occurrence of flavor conversions above the neutrino emission surfaces
in our merger remnant models in Sec.~\ref{sec:results}. 

\subsection{Dispersion relation formalism}
\label{sec:dispersion}
We adopt the density matrix formalism to describe the statistical properties of 
the neutrino dense gas incorporating flavor mixing~\cite{Sigl:1992fn}.
For a given density matrix $\R(\bm p,\bm x, t)$,
its diagonal elements in the flavor basis, $\R_{\alpha\alpha}$, record the phase-space distributions $f_{\nu_\alpha}$
of a given neutrino flavor $\nu_\alpha$ at the space-time location $(t,\bm x)$ and with momentum $\bm p$. 
The off-diagonal terms $\R_{\alpha\beta}$ carry the information about the 
neutrino mixing (neutrino flavor correlations). 
In the absence of any mixing, i.e., all neutrinos are in their flavor eigenstates, 
the off-diagonal elements vanish.

Neglecting general-relativistic effects and collisions of neutrinos with matter, 
the space-time evolution of the density matrix $\R(\bm p,\bm x, t)$ 
is governed by a Liouville equation
\begin{equation}
	\partial_t\R(\bm p,\bm x, t) + \bm {v_p}\cdot\nabla\R(\bm p,\bm x, t) = -i[\W(\bm p,\bm x, t),\R(\bm p,\bm x, t)],
	\label{flav_Lioville}
\end{equation}
where $\W$ is the Hamiltonian that accounts for the flavor oscillations of neutrinos. 
On the left hand side of Eq.~(\ref{flav_Lioville}), the first term takes care of the explicit 
time dependence of $\R$ and the second term takes into account the neutrino propagation 
with velocity $\bm{v_p}\simeq {\bm p}/|\bm p|$ for ultrarelativistic neutrinos. 
The Hamiltonian matrix $\W$ on the right hand side can be decomposed as
\begin{equation}
	\W(\bm p,\bm x, t) = \W_{\text{vac}} + \W_{\text{MSW}} + \W_{\nu\nu},
	\label{hamiltonian}
\end{equation}
where the first term $\W_{\text{vac}}$ takes into account flavor conversions in vacuum. 
In a simplified two-flavor scenario, $\W_{\text{vac}} = \text{diag}(\omega_{\rm v}/2, -\omega_{\rm v}/2)$ 
in the mass basis with $\omega_{\rm v} = (m_2^2 - m_1^2)/2E$ being the vacuum oscillation 
frequency of the neutrinos with energy $E$. 

The second term on the right hand side of Eq.~(\ref{hamiltonian}) embodies 
the effects of neutrino coherent forward scattering with electrons and nucleons. 
In the flavor basis, this term can be expressed as
\begin{equation}
    \W_{\text{MSW}} = (\sqrt{2}\Gf n_e) \text{diag}(1,0),
    \label{Hmsw}
\end{equation}
where $n_e$ is the 
net electron number density. 
The last term in Eq.~(\ref{hamiltonian}) is the effective Hamiltonian 
due to the $\nu$--$\nu$ interaction. 
For a neutrino traveling with momentum $\bm p$, $\W_{\nu\nu}$ is given by
\begin{equation}
    \W_{\nu\nu} = \sqrt{2}\Gf\int\frac{d^3\bm{q}}{(2\pi^3)}(1-\bm{v_p\cdot v_q})(\R(\bm q,\bm x, t) - \bar\R(\bm q,\bm x, t)),
    \label{Hnunu}
\end{equation}
where $\bar\R$ is the corresponding density matrix for antineutrinos.
The presence of $(1-\bm{v_p\cdot v_q})$ in Eq.~(\ref{Hnunu}) leads to 
multi-angle effects, i.e., neutrinos propagating 
in different directions experience different $\W_{\nu\nu}$. 
The equation of motion for antineutrinos can be obtained in a similar fashion, by  replacing $\omega_{\rm v}$ by $-\omega_{\rm v}$ in $\Omega_{\rm vac}$.

We focus on a simplified system that deals with two neutrino flavors.
Under this assumption, both the density matrix $\R$ and the Hamiltonian $\W$ 
are $2 \times 2$ Hermitian matrices and hence can be expanded in terms of the identity matrix 
and three Pauli matrices. 
Thus, we write $\R  = [(f_{\nu_e} + f_{\nu_x}) + (f_{\nu_e} - f_{\nu_x})\xi]/2$ for neutrinos 
and $\bar\R  = -[(f_{\bar\nu_e} + f_{\bar\nu_x}) + (f_{\bar\nu_e} - f_{\bar\nu_x})\xi^*]/2$ for antineutrinos.
The entity $\xi$ is a matrix defined as
\begin{equation}
    \xi = \begin{pmatrix}
          s && S\\
          S^* && -s
          \end{pmatrix},
\end{equation}
where $-1\le s \le 1$ and $|s|^2 + |S|^2$ =1. In the absence of any flavor correlation $S = 0$. 
Furthermore, as in previous work that studied the fast neutrino flavor conversion, 
we omit the vacuum oscillation term in the following discussion as $\omega_{\rm v}$ marginally affects the linear regime~\cite{Chakraborty:2016lct,Airen:2018nvp,Shalgar:2020xns}. 
With these assumptions and introducing the metric tensor $\eta^{\mu\nu} = \tx{diag}(1,~-1,~-1,~-1)$ 
and for any contra-variant vector $A^\mu$, $A_\mu = \eta_{\mu\nu}A^\nu$, 
we can recast the Hamiltonian defined in Eq.~(\ref{hamiltonian}) into the following form
\begin{equation}
    \W = v^\mu\lambda_\mu \frac{\sigma_3}{2}+ \int d\bm\Gamma' v^\mu v_\mu' \xi(\bm v') g(\bm v'),
    \label{covariant_hamiltonian}
\end{equation}
where $\bm v = (\sin\theta\cos\phi,~ \sin\theta\sin\phi, ~\cos\theta)$ 
with velocity  $v^\mu = (1,\bm v)$, $d\bm\Gamma = \sin\theta d\theta d\phi$
and $\lambda^\mu = \sqrt{2}\Gf n_e (1, \bm v_{\tx{m}})$ with $\bm v_{\tx{m}}$ 
being the vector  of the fluid  velocity of the background matter. 
Since the rate of pairwise conversion is much faster than any other inverse time scale involved in the problem, 
we treat the background matter as stationary and homogeneous:
 $\lambda_\mu v^\mu = \lambda_0$. 

The quantity $g(\bm v)$ is related to the angular distribution of the neutrino ELN angular distribution
\begin{equation}
    g(\bm v ) = \sqrt{2} \Gf (\bm\Phi_{\nu_e} - \bm\Phi_{\bar\nu_e}),
\end{equation}
where $\bm\Phi_{\nu_\alpha} = dn_{\nu_\alpha}/d\bm\Gamma$.
To study the growth of $S$ in the linear regime, 
we treat the flavor correlation $S$ 
as a perturbation and neglect all terms of $O(S^2)$ or higher. 
Taking $S(t,\bm x) = Q(\w,\bm k)e^{-i(\w t - \bm k \cdot \bm x )}$,
the EoM becomes
\begin{equation}
    v_\mu\bar\lambda^\mu Q(\w,\bm k) = -\int d\bm\Gamma'v^\mu v_\mu'g(\bm v')Q(\w,\bm k).
    \label{eom_fourier}
\end{equation}
In the above Eq.~(\ref{eom_fourier}) we have introduced the four 
vector $\bar\lambda^\mu = (\w - \lambda_0- \epsilon_0,\bm k - \bm\epsilon)$, 
$\epsilon_0 \equiv \int d\bm\Gamma g(\bm v)$ and $\bm\epsilon\equiv\int d\bm\Gamma \bm v g(\bm v)$. 
Inspecting Eq.~(\ref{eom_fourier}), one can make the ansatz 
$Q(\w, \bm k) =  v_\mu a^\mu/v_\mu\bar\lambda^\mu$, 
with $a^\mu$ being the coefficients of eigenfunction solutions. 
Thus, Eq.~(\ref{eom_fourier}) becomes
\begin{equation}
    v_\mu\Pi^{\mu\nu}a_\nu = 0,
    \label{eom_Pi}
\end{equation}
where we have used the definition
\begin{equation}
    \Pi^{\mu\nu} \equiv \eta^{\mu\nu}  + \int d\bm\Gamma g(\bm v) \frac{v^\mu v^\nu}{v_\sigma\bar\lambda^\sigma}.
    \label{Pi}
\end{equation}
The EoM defined in Eq.~(\ref{eom_Pi}) holds for any $v^\mu$. 
Thus, we have the condition $\Pi^{\mu\nu}a_\nu = 0$. 
Eigenfunctions of the latter have non-trivial solution only if $\Pi^{\mu\nu}$ satisfies the condition
\begin{equation}
    \tx{det}[\Pi^{\mu\nu}(\w, \bm k)] = 0.
    \label{dispersion}
\end{equation}
Equation~(\ref{dispersion}) is the DR in flavor space. 
The solutions of the DR have been classified into several types~\cite{Capozzi:2017gqd,Yi:2019hrp}.
If $\w$ is real for real values of $\bm k$, 
a perturbation in $S$ only propagates without growing or damping, 
i.e., it stays in the linear regime, meaning that no significant flavor conversion occurs. 
On the other hand, an imaginary solution of $\w$ with $\tx{Im}(\w)>0$ 
corresponds to exponentially growing modes. 
In other words, the flavor correlation $|S|$ grows exponentially with time, 
leading to significant flavor conversion. 
Rigorous studies have been carried out to understand the characteristics of 
the above DR with respect to the ELN angular distribution of  neutrinos~\cite{Izaguirre:2016gsx,Capozzi:2017gqd,Yi:2019hrp}. 
It was shown that in the presence of a crossing in the ELN distribution, 
the DR will yield complex $\w$ solutions for real $\bm k$, 
leading to temporal instabilities. 
In the following, we examine the ELN distributions above the merger remnants
and the corresponding flavor instabilities.

\subsection{Neutrino electron lepton number angular distribution}
\label{sec:ELN}

\begin{figure*}[ht]
\includegraphics[width = \textwidth]{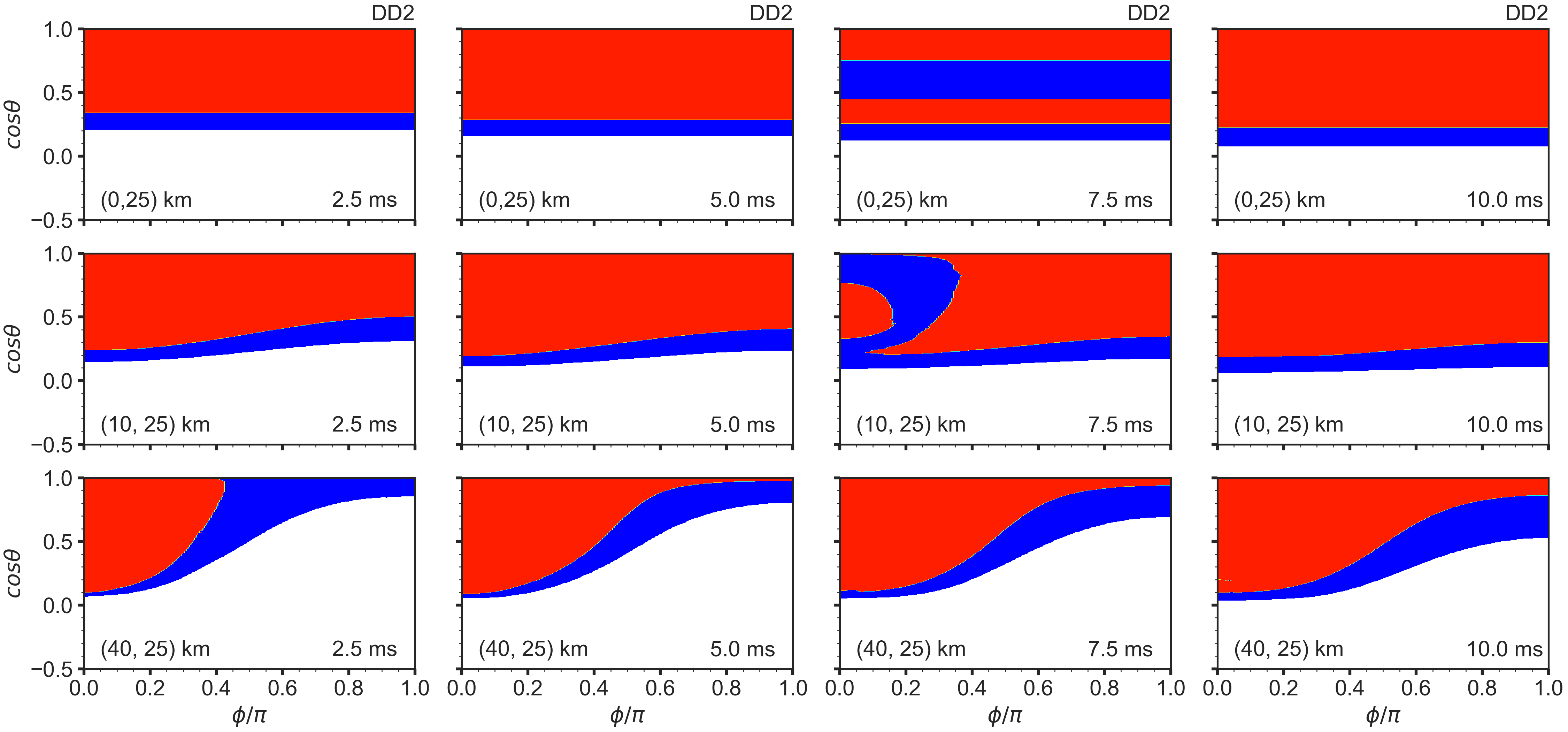}
\caption{Neutrino electron lepton number angular crossings at different positions ($x,z)=(0,25),(10,25),(40,25)$~km and times $t=2.5, 5, 7.5, 10$~ms, above the remnant for the $1.35+1.35$~$M_\odot$ model with DD2 EoS. 
The red (blue) shaded region represents the region where $\Phi_{\nu_e} - \Phi_{\bar\nu_e} > 0$ ($\Phi_{\nu_e} - \Phi_{\bar\nu_e} < 0$). 
}
\label{fig:eln}
\end{figure*}
\begin{figure*}[ht]
\includegraphics[width = \textwidth]{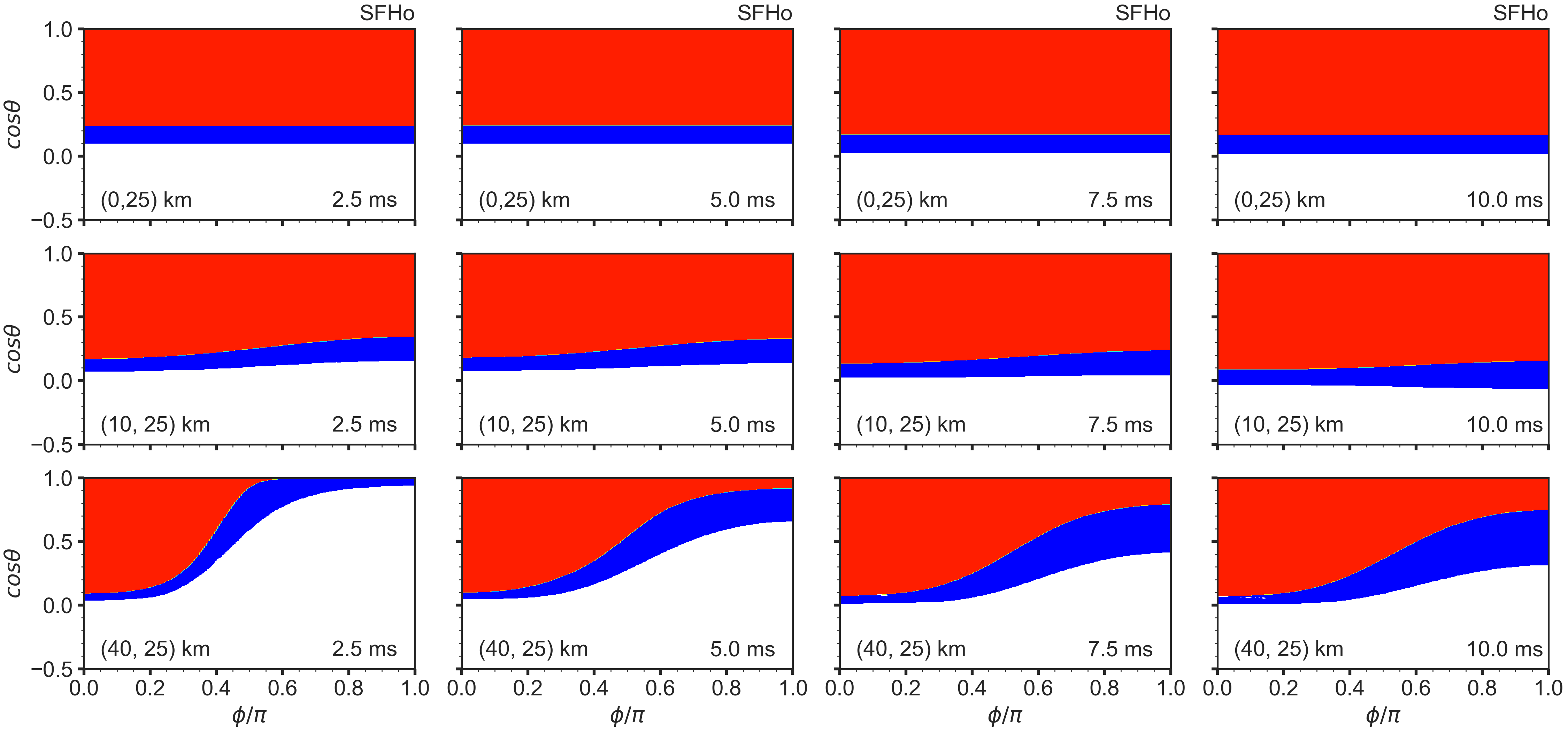}
\caption{Same as Fig.~\ref{fig:eln}, but for the $1.35+1.35$~$M_\odot$ simulation with SFHo EoS. }
\label{fig:eln2}
\end{figure*}

To construct the ELN distribution above the merger remnant, 
we follow a  method similar to the one adopted in Ref.~\cite{Wu:2017drk}. 
First, we assume that both $\nu_e$ and $\bar\nu_e$ freely 
propagate outside their respective emission surfaces defined 
in Sec.~\ref{sec:nu_surf}.
Second, we approximate their forward-peaked angular distributions at each point
on the emission surfaces as
\begin{equation}
    {\bm\Phi}_{\nu_e, \bar\nu_e}(\theta_n) = \frac{n_{\nu_e, \bar\nu_e}}{4\pi}(1+\cos\theta_n),
    \label{emission_spectrum}
\end{equation}
where $\theta_n$ is the angle with respect to the normal direction 
of the location on the emission surface. 
Ignoring the minor effect of GR bending, we can then ray-trace
the neutrino intensities from the emission surfaces
to obtain their angular distributions at any location above the
surfaces.

Figures~\ref{fig:eln} and \ref{fig:eln2} show the obtained ELN distributions as a function of 
the local angular variables $\theta$ (angle with respect to the $z$-axis) 
and $\phi$ (angle with respect to the $x$-axis on the $x$-$y$ plane) at selected locations
above the $\nu_e$ surface at 2.5, 5, 7.5 and 10~ms after the coalescence, for the $1.35+1.35$~$M_\odot$ merger models with DD2 and SFHo EoS, respectively. 
Here we only show the sign of the ELN distribution to highlight the crossing.
The blue shade represents the region where the net ELN is positive 
while the red region corresponds to negative ELN. 
The top panel shows the ELN distribution at a representative 
point on the $z$ axis and the middle and bottom panels show the 
ELN distributions at near-center and outer regions above the $\nu_e$ surface. 
Note here that Figs.~\ref{fig:eln} and \ref{fig:eln2} only show the ELN distribution for 
$\phi/\pi \ge 0$ as the distributions for axi-symmetric emission surfaces possess reflection symmetry, 
$\Phi_{\nu_e, \bar\nu_e}(\cos\theta,\phi) = \Phi_{\nu_e,\bar\nu_e}(\cos\theta,-\phi)$.

The angular coverage of the $\nu_e$ and $\bar\nu_e$ fluxes at a point 
($x,z$) is determined by the geometry of the $\nu_e$ and $\bar\nu_e$ emission surfaces. 
On the other hand, the ELN angular distribution is determined by 
the combination of the emission surface geometries and their 
respective emission properties, 
including the relative strength of $\nu_e$ and $\bar\nu_e$ and the angular dependence of the emission.
For example, the ELN angular distribution at any point on the $z$ axis 
above the emission surface is independent of $\phi$,
reflecting the (assumed) rotational symmetry of the merger remnants about the $z$ axis. 
As we move away from the $z$ axis, the ELN distribution becomes dependent on $\phi$ (second and third panels). 
For both EoS, the angular coverage in $\theta$ slightly increases 
with time for a given $(x,z)$, 
caused by the expansion of the emission surfaces. 
This is different from what was found in Ref.~\cite{Wu:2017drk} 
where the central remnant is a BH for which the size of the
$\nu_e/\bar\nu_e$ emitting torus surfaces shrinks with time. 
As $\bar\nu_e$ are more abundant than $\nu_e$ in most parts of the emission region during the first
10~ms (see Fig.~\ref{fig:neudensities}),
the overall shapes of the ELN crossings are qualitatively similar.
The only exception is represented by the snapshot at 7.5~ms for the case with DD2 EoS, for which
the inner region above the merger remnant shows
a double ELN crossing structure, see the left and middle panels in
the third row.
This is related to the dip of the $\bar\nu_e$ density
on the $\bar\nu_e$ surface at $\sim 10$~km discussed in 
Sec.~\ref{sec:nu_surf}. Also, the ELN angular distribution in Fig.~\ref{fig:eln2} is more extended toward $\theta = 0$ compared to Fig.~\ref{fig:eln}, resulting from slightly larger radii of the neutrino emission surfaces in the SFHo model.
For the simulations adopting $1.25+1.45$~$M_\odot$ binaries, we have similarly checked the ELN distributions during the same time snapshots for both EoS. Unsurprisingly, ELN crossings appear at all times as those shown here.

\subsection{Flavor instabilities for fast pairwise conversion}
\label{sec:results}
\begin{figure*}[h]
\includegraphics[width = \linewidth]{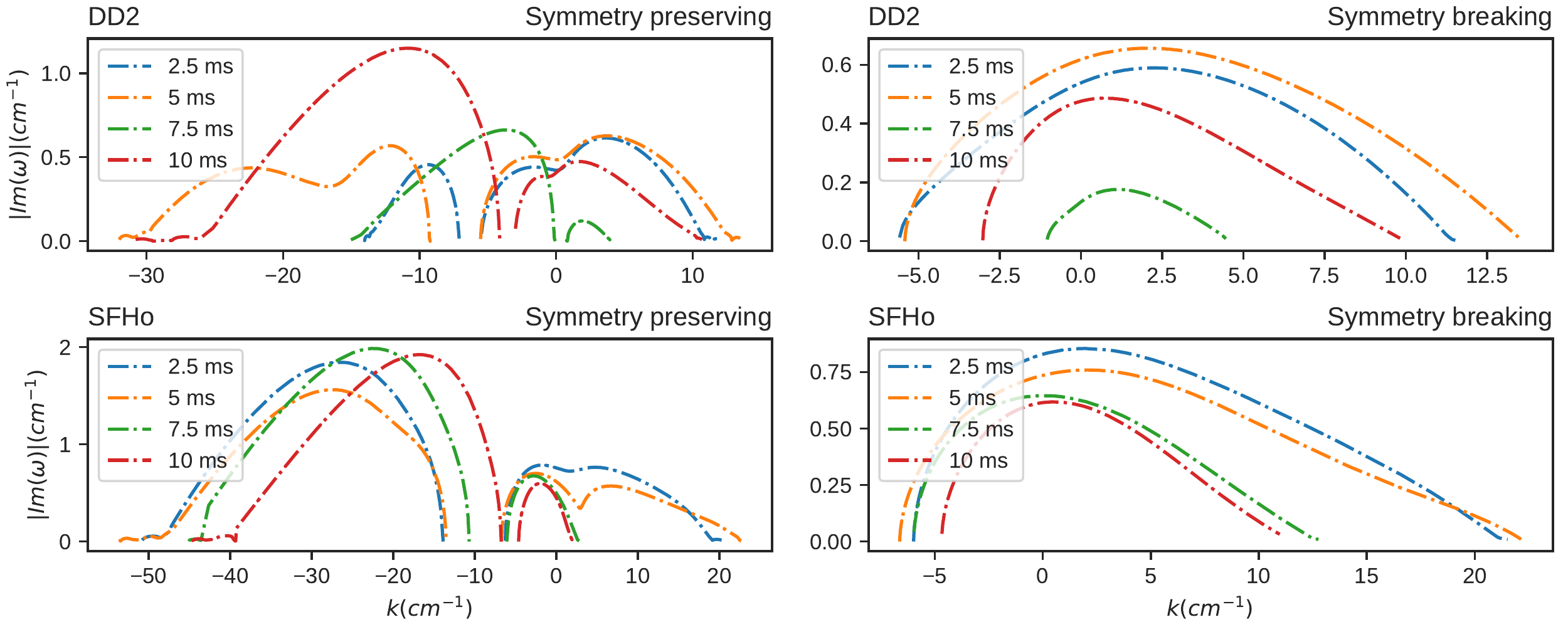}
\caption{
Solutions of $|\tx{Im}(\w)|$ obtained from the DR [Eq.~(\ref{dispersion})] for ${\bm k}=(0,0,k)$ at the location $(x,z)=(10,25)$~km above the merger remnant for different times.
The top (bottom) panels show the results for the $1.35+1.35$~$M_\odot$ model with the DD2 (SFHo) EoS, while the left (right) panels show the symmetry-preserving (symmetry-breaking) solutions. 
}
\label{fig:Wi}
\end{figure*}
\begin{figure*}
        \includegraphics[width = \linewidth]{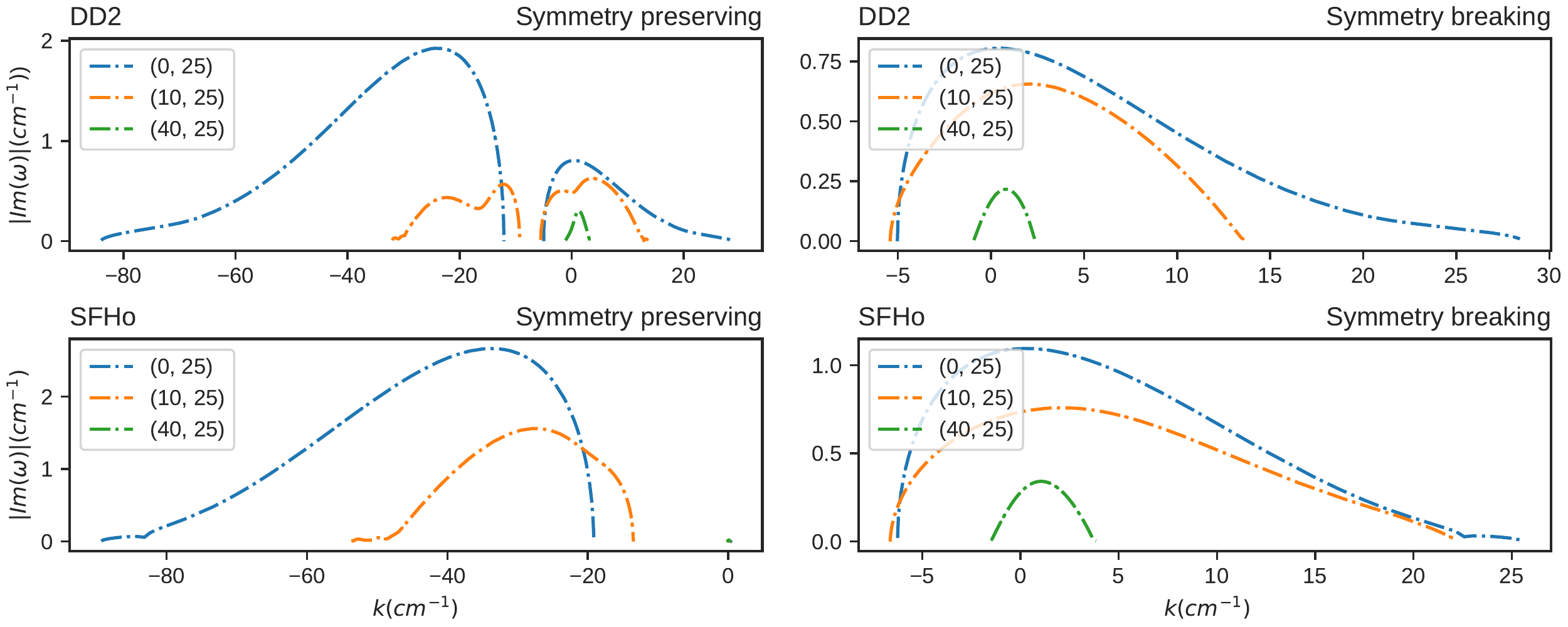}
\caption{
Solutions of $|\tx{Im}(\w)|$ obtained from the DR [Eq.~(\ref{dispersion})] for ${\bm k}=(0,0,k)$ at $5$~ms for different locations labeled by the coordinates $(x,z)$ expressed in km. The top (bottom) panels show the results for the $1.35+1.35$~$M_\odot$ model with the DD2 (SFHo) EoS, while the left (right) panels show the symmetry-preserving (symmetry-breaking)  solutions.
}
\label{fig:Wi_2}
\end{figure*}

After obtaining the ELN distributions above the neutrino emitting surface, we  numerically solve the 
DR [Eq.~\eqref{dispersion}] starting from the outer neutrino surface to inspect whether solutions containing
non-zero $\rm{Im}(\omega)$ for a given $\bm k$ can be found\footnote{Note that the positive and negative $\rm{Im}(\omega)$ solutions always appear together.}.
As the ELN distributions above the merger remnants with axial-symmetry
have a reflection symmetry with respect to $\phi\rightarrow -\phi$, 
one can obtain two different solutions that correspond to
the reflection symmetry-preserving and symmetry-breaking cases~\cite{Wu:2017qpc}.

We show the obtained $|\tx{Im}(\w)|$ as a function of $k_z$, taking $k_x=k_y=0$, 
at different times for a location at $(x,z)=(10,25)$~km above the emitting surfaces in Fig.~\ref{fig:Wi} and
the solutions for different locations corresponding to the ELN crossing 
shown in Fig.~\ref{fig:Wi_2} at 5~ms.
The top (bottom) panels are for cases with DD2 (SFHo) EoS while
the left (right) panels show the symmetry-preserving (symmetry-breaking) solutions. 

Flavor instabilities with growth rate of $\mathcal{O}(1)$~cm$^{-1}$ 
exist at all locations at all times for a large range of $k_z$.
The symmetry preserving solution of the DR has two branches 
while the symmetry breaking solution has only one branch, 
similar to what was found in Ref.~\cite{Wu:2017qpc}. 
Comparing the results for the models with DD2 and SHFo EoS, 
the growth rate of the flavor instability is generally larger in the latter,
as the neutrino emission is stronger with the SHFo EoS.
The shape of the solutions is rather stable over time for the case with SFHo EoS. 
On the other hand, the model with DD2 EoS shows a somewhat stronger time-dependence.
In particular, the range of $k_z$ that leads to non-zero $|{\rm Im}(\omega)|$
as well as the value of $|{\rm Im}(\omega)|$ both decrease at the snapshot of 7.5~ms, 
when the double crossing shape of the ELN appears (see Fig.~\ref{fig:eln}).

By comparing the solutions at different locations for the $t=5$~ms snapshot, Fig.~\ref{fig:Wi_2} shows that  $|{\rm Im}(\omega)|$ is larger  
closer to the $z$-axis for both the symmetry-preserving and 
symmetry-breaking solutions. 
This, again, is caused by the fact that
$\Phi_{\nu_e}$ and 
$\Phi_{\bar\nu_e}$ are largest in this region (see Fig.~\ref{fig:neudensities}). 

\begin{figure*}[ht]
\begin{centering}
\includegraphics[width = \linewidth]{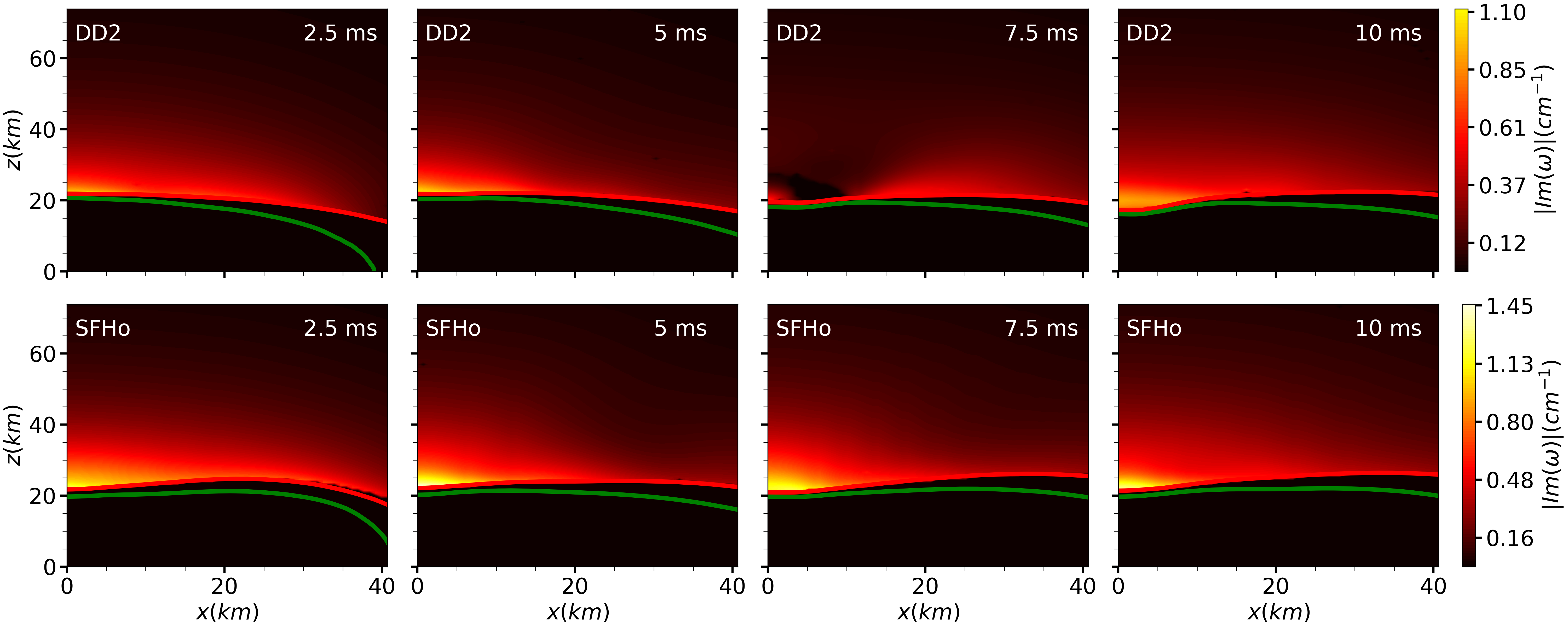}
\end{centering}
\caption{Contour plots showing $|\tx{Im}(\w)|$ above the $\nu_e$ surface for $\bm{k}=0$ at $t=2.5$, $5.0$, $7.5$ and $10.0$~ms,  from left to right, for the $1.35+1.35$~$M_\odot$ model. 
The top (bottom) panels are based on the model with the DD2 (SFHo) EoS. 
The red and green solid lines represent the locations of $\bar\nu_e$ and $\nu_e$ emission surfaces, respectively.
}  
\label{fig:colormap}
\end{figure*}

As shown in Figs.~\ref{fig:Wi} and \ref{fig:Wi_2}, the symmetry-breaking 
solutions at all times and all locations contain non-zero $|\tx{Im}(\w)|$
for the mode $\bm k=0$.
We further show in Fig.~\ref{fig:colormap} the contour plot of $|\tx{Im}(\w)|$  
above the neutrino surface for this mode. 
The plots in this figure clearly show that growth rates of the instability of $\sim {\cal O}(1)$ cm$^{-1}$ exist everywhere above the remnant at all times.
The region closer to the $z$-axis just above the emission surface has the maximal growth rates. 
Moving away from the surface, the neutrino flux is suppressed by the geometric effects and the magnitude of the instability decreases in all  cases.
For the sake of comparison with the existing literature, we find $|\rm{Im}(\omega)/\mu|\sim\mathcal{O}(10^{-2})$ for all cases examined here;
this value of the growth rate is similar albeit a bit smaller than the ones reported in Refs.~\cite{Wu:2017drk,1815221}. The growth rates $|\rm{Im}(\omega)/\mu|$ are generally larger toward the middle region above the $\nu_e$ surface, in agreement with the findings of
Ref.~\cite{Wu:2017qpc,Wu:2017drk}.
This is due to the relative strength of the positive vs.~negative ELN strength, in the proximity of crossings between the $\nu_e$ and $\bar\nu_e$ angular distributions~\cite{Shalgar:2019qwg}.
For a system that is more dominated by $\nu_e$ or $\bar\nu_e$, i.e., where the positive ELN distribution dominates the negative parts or the other way around, one expects that 
the value of $|\rm{Im}(\omega)/\mu|$ is smaller than in a more balanced system with similar positive and negative parts of the ELN distribution (see e.g., Ref.~\cite{Yi:2019hrp}).
Looking at the ELN crossing pattern shown in 
Figs.~\ref{fig:eln} and \ref{fig:eln2},
the locations above the middle part of the remnant
have large angular area of $g(\bm{v})>0$ 
due to the larger separation of the $\nu_e$ and $\bar\nu_e$ surfaces (see Fig.~\ref{fig:simsnapDD2} and \ref{fig:simsnapSFHo}). 
This, in turn, enhances the corresponding values of $|\rm{Im}(\omega)/\mu|$ relative to the ones in the inner region closer to the $z$-axis.

Likewise, when comparing the values of  $|\rm{Im}(\omega)/\mu|$ to the ones shown in Ref.~\cite{Wu:2017drk}, the more dominant $\bar\nu_e$ emission relative
to $\nu_e$ here, together with the smaller separation of their emission surfaces,
results in smaller values of $|\rm{Im}(\omega)/\mu|$.

On the other hand, while Ref.~\cite{Wu:2017drk} showed that the
region where the flavor instability exists shrinks on a time scale of $\sim \mathcal{O}(10)$~ms as the BH-disk remnant evolves, 
the instability region found here remains stable within the examined 10~ms of post-merger evolution.
This can have important consequences for the 
growth of the instabilities and seems to favor the 
eventual development of flavor conversions in the non-linear regime. 
The effect of advection hindering the growth of the 
flavor instabilities for systems with non-sustained and  fluctuating unstable conditions shown by Ref.~\cite{Shalgar:2019qwg} may therefore not happen above the merger remnant, as shown in Ref.~\cite{1815221}.
In the case of the models with unequal mass binaries, due to the similar ELN crossing features, we find  unstable regions above the merger remnant disk and growth rates very similar to the symmetric models (results not shown here).

\begin{figure*}[ht]
    \includegraphics[width =1.05 \linewidth]{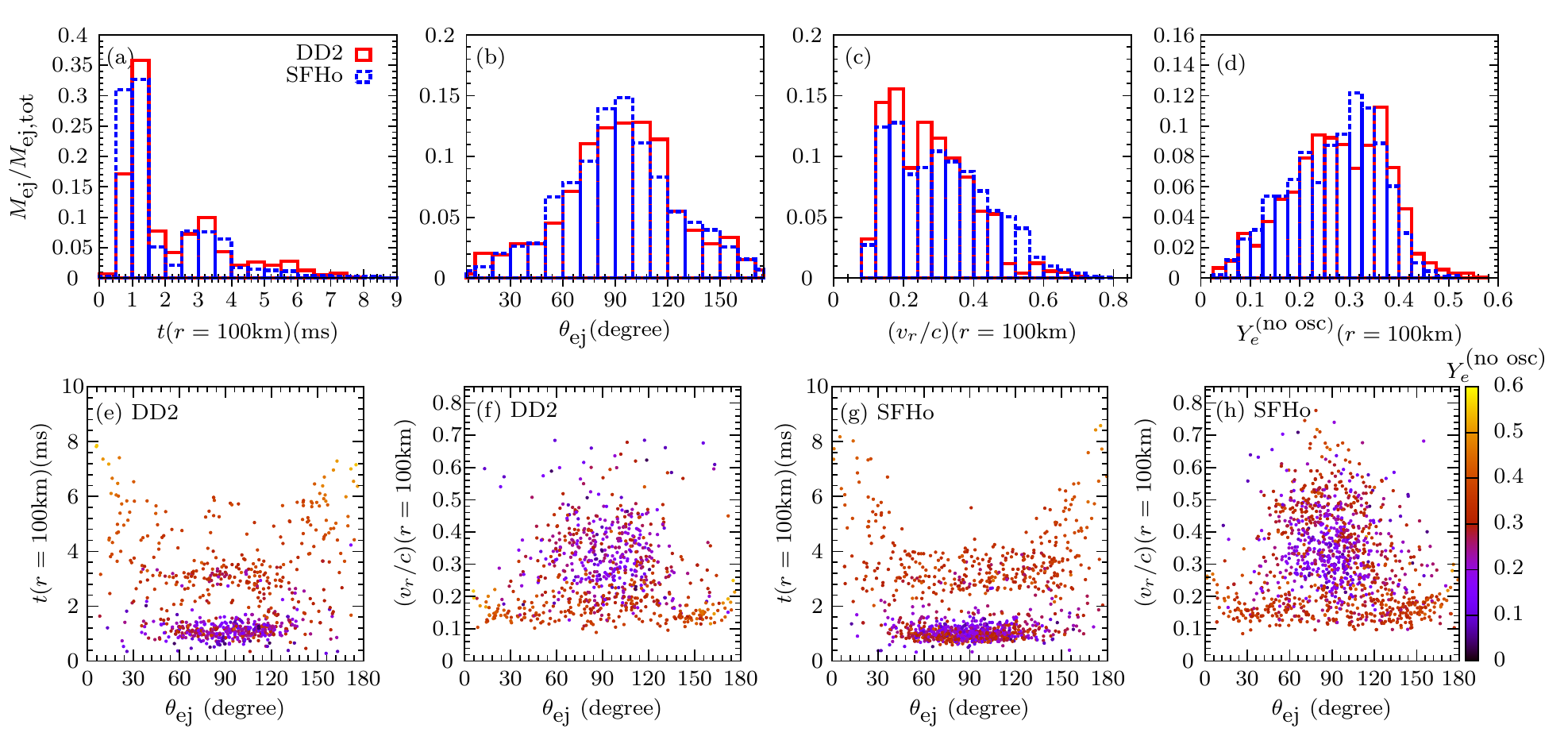}
    \caption{Histograms of the ejecta mass fraction when the ejecta reach $r=100$~km at $t(r=100~{\rm km})$, [panel (a)], $\theta_{\rm ej}$ [panel (b)], radial velocity $v_r$ at $r=100$~km, and $Y_e$ at $r=100$~km without flavor conversions for the simulations with $1.35+1.35$~$M_\odot$ NS binaries.
    Panel (e) [(g)] and (f) [(h)] show the distributions of each tracer particle in the $t(r=100)$~km--$\theta_{\rm ej}$ and $(v_r/c)$--$\theta_{\rm ej}$ planes with the DD2 (SFHo) EoS. The color coding of each tracer particle indicates its $Y_e$ value at $r=100$~km without considering flavor conversions.
    }
    \label{fig:ejpropsym}
\end{figure*}

We here focus on the diagnostics of  flavor instabilities, but do not distinguish whether they belong to the convective or absolute type (see Refs.~\cite{Capozzi:2018clo,Yi:2019hrp}). This would 
require solving the full dispersion relation 
for the complete $(\omega,\mathbf{k})$ space.

\section{Merger ejecta and nucleosynthesis of the heavy elements}
\label{sec:nucleo}

In this section, we first analyze the properties of the material ejected during the first $\simeq 10$~ms post merger, examine how neutrino absorption affects the evolution of $Y_e$  of the outflow material, and
discuss the nucleosynthesis outcome in the absence of neutrino flavor conversion in Sec.~\ref{sec:nuabsorb}.
In Sec.~\ref{sec:equiparti}, we further explore the potential effect of fast flavor conversion on the neutrino absorption rates and the outcome of nucleosynthesis of in these ejecta.

\subsection{Ejecta properties and nucleosynthesis}
\label{sec:nuabsorb}

The dynamical ejecta masses extracted at 10~ms post-merger for the $1.35+1.35$~$M_\odot$ merger simulations 
are $\sim 2.0\times 10^{-3}$~$M_\odot$ and $3.3\times 10^{-3}$~$M_\odot$ with the DD2 and SFHo EoS, represented by 783 and 1263 tracer particles, respectively.
For the $1.25+1.45$~$M_\odot$ merger cases, the dynamical ejecta masses are $\sim 3.2\times 10^{-3}$~$M_\odot$ (DD2) and $8.7\times 10^{-3}$~$M_\odot$ (SFHo), represented by 1290 and 4398 tracer particles.
In computing the evolution of $Y_e$ for a given tracer particle, we first post-process the neutrino emission data from Ref.~\cite{Ardevol-Pulpillo:2018btx} to calculate the absorption
rates of $\nu_e$ and $\bar\nu_e$ on neutrons and protons, $\lambda^0_{\nu_e}$ and $\lambda^0_{\bar\nu_e}$ (the superscript $0$ here denotes the case where any neutrino flavor conversion is omitted), as detailed in Appendices~\ref{app:postproc} and \ref{app:rates}.
These rates are then combined with the nuclear reaction network used in Refs.~\cite{Wu:2016pnw,Wu:2018mvg} to compute the nucleosynthesis yields in the merger ejecta.
For each tracer particle, we begin the network calculation either at  a location where $T=50$~GK or just outside the disk with  height of 25~km and  radius of 55~km to make sure that it is outside the $\nu_e$ emission surface. 
The initial nuclear abundances are calculated by using the nuclear statistical equilibrium (NSE) condition. For $T>10$~GK, we only compute the weak reaction rates to track the $Y_e$ evolution while assuming NSE at each moment to obtain the abundances. 
When $T\leq10$~GK, we instead follow the full evolution of all nuclear species and include the feedback due to the nuclear energy release on the ejecta temperature following Ref.~\cite{Mendoza-Temis:2014mja}.

In Fig.~\ref{fig:ejpropsym}, we show the distributions of the ejecta mass for the $1.35+1.35$~$M_\odot$ models as a function of the time when the ejecta reach the radius  $r=100$~km at $t(r=100~\rm{km})$ in panel (a), the angle $\theta_{\rm ej}$ (relative to the $z$-axis) when they leave the simulation domain in panel (b), and the radial velocity $v_r$ at $r=100$~km in panel (c).
We also show, in panel (d), the distributions of $Y_e$ for the ejecta  at $r=100$~km, in the absence of flavor mixing. 
Additionally, panels (e) and (g) [(f) and (h)] show the distributions of each tracer particle in the plane spanned by $t(r=100~\rm{km})$ [$v_r(r=100)~\rm{km}$] and $\theta_{\rm ej}$ with the corresponding $Y_e$ at 100~km labeled in color for the DD2 and SHFo EoS, respectively.
These plots show that, independently of the adopted EoS, most of the material is ejected within the first $\sim 2$~ms with lower $Y_e\lesssim 0.3$, within $\sim 45^\circ$ away from the mid-plane. 
The ejecta launched in the second episode [$t(r=100~\rm{km})\simeq 3-4$~ms] are similarly distributed closer to the equatorial plane and have $Y_e\simeq 0.4$.
After that, neutrino irradiation continues to power the mass ejection along the polar direction with $Y_e$ reaching $\sim 0.5$.
In terms of the ejecta kinematics, because of the weak interactions, it is clear that $Y_e$ can  be raised up to $0.3-0.4$ mostly for the material expanding more slowly with $v_r/c\sim 0.1-0.2$ at $r=100$~km, even around the equatorial plane,  and reach $\sim 0.5$ closer to the polar direction.

\begin{figure*}[ht]
    \includegraphics[width = 1.05 \linewidth]{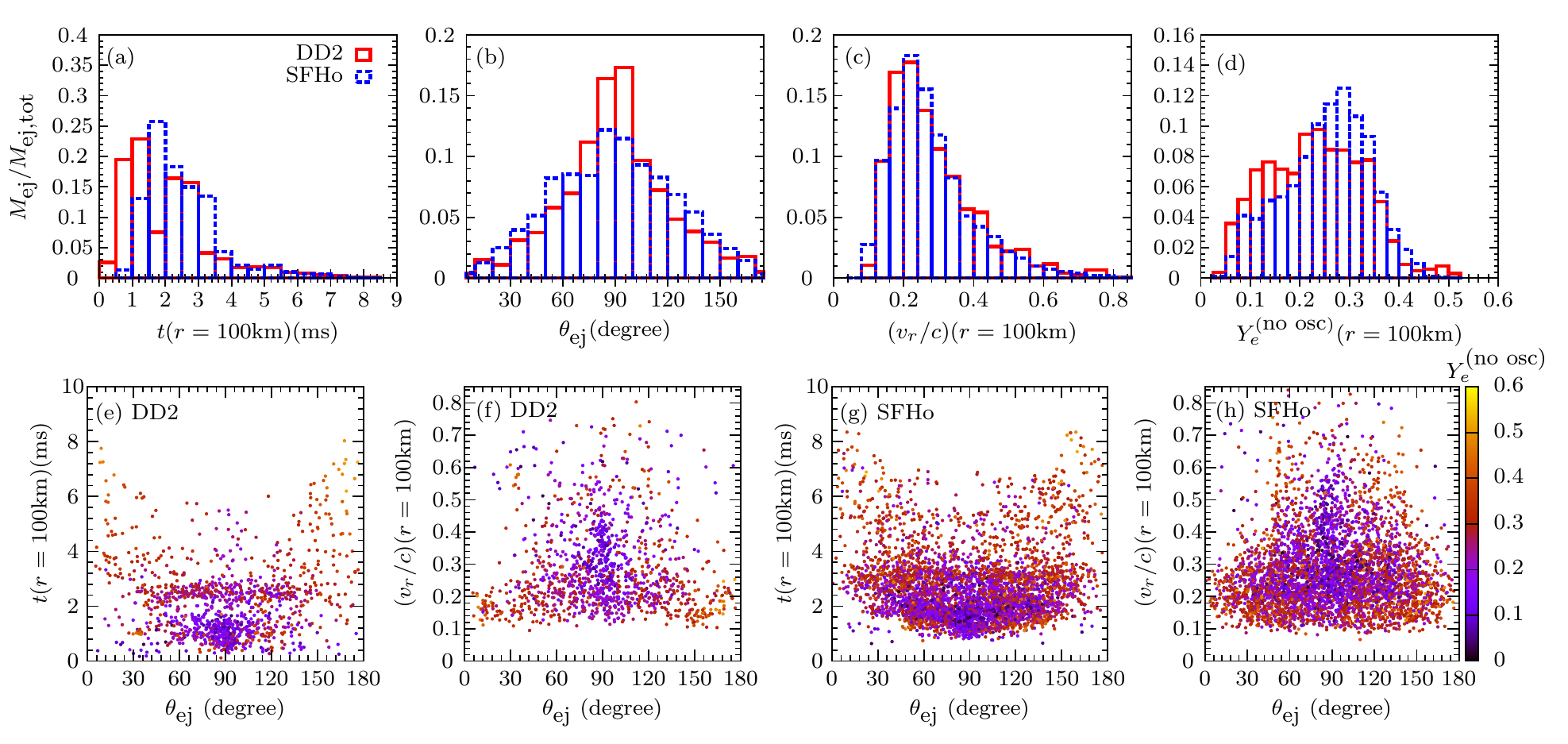}
    \caption{Same as Fig.~\ref{fig:ejpropsym}, but for the   $1.25+1.45$~$M_\odot$ models.}
    \label{fig:ejpropasym}
\end{figure*}

The main differences between the  DD2 and the SFHo EoS results are  the following. 
First, for the simulation with SHFo EoS, the amount of ejecta is larger, particularly during the first ejection episode [see panel (a)], than that with the DD2 EoS;
this is a direct consequence of the more violent collision of the two NSs in the case with softer EoS  (SHFo), as discussed in Ref.~\cite{Bauswein:2013yna}. 
Second, the amount of high-velocity ejecta with $v_r/c\gtrsim 0.5$ for the SFHo case  is also higher for the same reason [see panels (c) and (f)].
Third, the $Y_e$ distribution [panel (d)] for both models shows that $Y_e(r=100~\mathrm{km})$ is slightly larger for the DD2 EoS model than for the SFHo EoS one. 
This reflects the higher $\bar\nu_e$ luminosity obtained in the SHFo EoS model (see Fig.~\ref{fig:lumegy}) leading to larger  $\bar\nu_e$ absorption rates on protons; hence,  $Y_e$ is raised less for the bulk of the ejecta. 
The high $Y_e$ tail extends to values $\gtrsim 0.5$ in the DD2 EoS model, while it remains  $\lesssim 0.5$ in the SFHo EoS model [see panels (c) and (e)].
Interestingly, it is noticeable that the high velocity components in the SFHo model have  
more high-$Y_e$ ejecta than in the DD2 model [see panel (f)],
different from the main bulk of the ejecta. 
This is because these ejecta are mainly driven during the early post-merger phase. 
During this early phase, positron capture is responsible for raising $Y_e$ in the ejecta, rather than neutrino absorption.
Thus, in the model with SFHo EoS, the higher post-merger temperature of the remnant due to the more violent merger leads to relatively higher $Y_e$ material in the early fast ejecta.

We show in Fig.~\ref{fig:ejpropasym} the same quantities as in Fig.~\ref{fig:ejpropsym}, but for the models  with $1.25+1.45$~$M_\odot$ mass binaries.
Qualitatively, the  features are similar to the ones described above, but they are quantitatively different. 
For instance, the division between different episodes of mass ejection is less clear, in particular for the  SHFo EoS model [see panel (a)]. 
On average, most of the ejecta have higher velocities, peaked at $\sim 0.25c$ [see panel (c), (f) and (h)].
Moreover, a larger fraction of the ejecta has lower $Y_e\lesssim 0.2$, despite the fact that similarly wide distributions of $0.01\lesssim Y_e\lesssim 0.5$ are obtained. 

\begin{figure}[ht]
    \includegraphics[width = \linewidth]{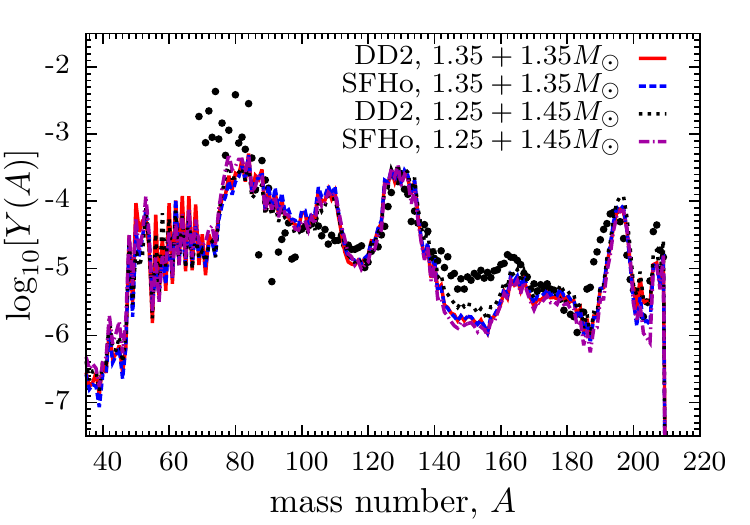}
    \caption{Nucleosynthesis yields, $Y(A)$, as a function of the nuclear mass number $A$ at the time of 1~Gyr for all NS merger models investigated in this work.}
    \label{fig:nucleoall}
\end{figure}

Figure~\ref{fig:nucleoall} shows the resulting abundance distributions for the ejecta of the symmetric and asymmetric merger models  with the DD2 and SFHo EoS, in comparison to the re-scaled solar $r$ abundance pattern~\cite{Cowan:1998nv}.
Due to the wide-spread $Y_e$ distribution ranging from $0.02-0.5$ in all models, all three $r$-process peaks at $A\simeq 90$, $130$, and $195$ are in relatively good agreement with the solar abundance pattern.
The difference of the high $Y_e$ component discussed above only results in  abundance variations at $A\simeq 50-80$.
In particular, we compute the amount of strontium present at the time of a day, which is of relevance  to the potential identification in the spectral analysis of the GW170817 kilonova~\cite{Watson:2019xjv}. 
The total amount of strontium is $\simeq 8.9\times 10^{-5}$~$M_\odot$ and $\simeq 3.9\times 10^{-5}$~$M_\odot$ for the  DD2 and SFHo EoS models with equal mass binaries.
For the asymmetric merger models, the corresponding amount of strontium is  $\simeq 2.66\times 10^{-4}$~$M_\odot$ and $3.25\times 10^{-4}$~$M_\odot$, respectively. These results are consistent with the findings of Ref.~\cite{Watson:2019xjv}.

References~\cite{Metzger:2014ila,Mendoza-Temis:2014mja,Goriely:2016cxi,Bovard:2017mvn} found that some merger ejecta have low $Y_e$ and fast expansion time scale to allow for a neutron-rich freeze-out during the $r$-process nucleosynthesis.
A potentially thin layer of ``neutron-skin'' at the outskirts of the ejecta may possibly power an early-time UV emission at $\sim$ hours post-merger due to the radioactive heating of neutron decay~\cite{Metzger:2014ila,Metzger:2016pju}.
We find that the amount of free neutrons at the end of the $r$-process, for equal-mass binaries, is $\simeq 6.2\times 10^{-6}$~$M_\odot$ and $6.6\times 10^{-6}$~$M_\odot$ with the DD2 and SFHo EoS, respectively. 
These numbers are roughly a factor of 10 smaller than what was found in Ref.~\cite{Mendoza-Temis:2014mja}, which analyzed simulation trajectories without including the weak interactions. 
The reduction of the amount of free neutrons at the end of the $r$-process is related to the high post-merger temperature effect raising $Y_e$ even for the early fast ejecta. 
For the unequal mass binaries, the corresponding amount of free neutrons is $\simeq 9.2\times 10^{-6}$~$M_\odot$ and $3.1\times 10^{-6}$~$M_\odot$ for the DD2 EoS model and  for the SFHo model.
A future (non)identification of this component may  shed light on the role of weak interactions in the post-merger environments.

\subsection{Impact of flavor equipartition on $Y_e$ and nucleosynthesis}\label{sec:equiparti}

Following previous work~\cite{Wu:2017drk,Xiong:2020ntn}, we assume that fast flavor conversions lead to conditions close to flavor equipartition for neutrinos and antineutrinos. The assumption of flavor equilibration is an extreme ansatz, especially in the light of the findings of Ref.~\cite{1815221}, which, however, relied on a simplified model of
a relic merger disk. 
Moreover, this simple assumption does not preserve the total electron neutrino lepton number, which is a strictly conserved quantity in the case of pairwise flavor conversions.
Nevertheless, our extreme assumption for the flavor ratio is useful to explore the
largest possible impact that flavor conversions might have on the 
nucleosynthesis of the heavy elements.
The corresponding neutrino absorption rates can be approximated by
\begin{align}
    \lambda^{\rm osc}_{\nu_e} &= \frac{1}{3}~\lambda^0_{\nu_e} + \frac{2}{3}~\lambda_{\nu_x},\\
    \lambda^{\rm osc}_{\bar\nu_e} &= \frac{1}{3}~\lambda^0_{\bar\nu_e} + \frac{2}{3}~\lambda_{\bar\nu_x},
\end{align}
where $\lambda_{\nu_x}$ and $\lambda_{\bar\nu_x}$ are the neutrino absorption rates on free nucleons assuming that all $\nu_x$ and $\bar\nu_x$ are converted to $\nu_e$ and $\bar\nu_e$, as detailed in Appendices~\ref{app:postproc} and \ref{app:rates}.
We then perform the same nucleosynthesis calculations as in Sec.~\ref{sec:nuabsorb} for all tracer particles in all the merger models by replacing $\lambda^0_{\nu_e}$ and $\lambda^0_{\bar\nu_e}$ by $\lambda^{\rm osc}_{\nu_e}$ and $\lambda^{\rm osc}_{\bar\nu_e}$.
Below, we only focus on the findings for the models with equal mass and different EoS because the results obtained in the unequal mass binaries are qualitatively the same, independent of the EoS.
\begin{figure*}[ht]
    \includegraphics[width = 0.8\linewidth]{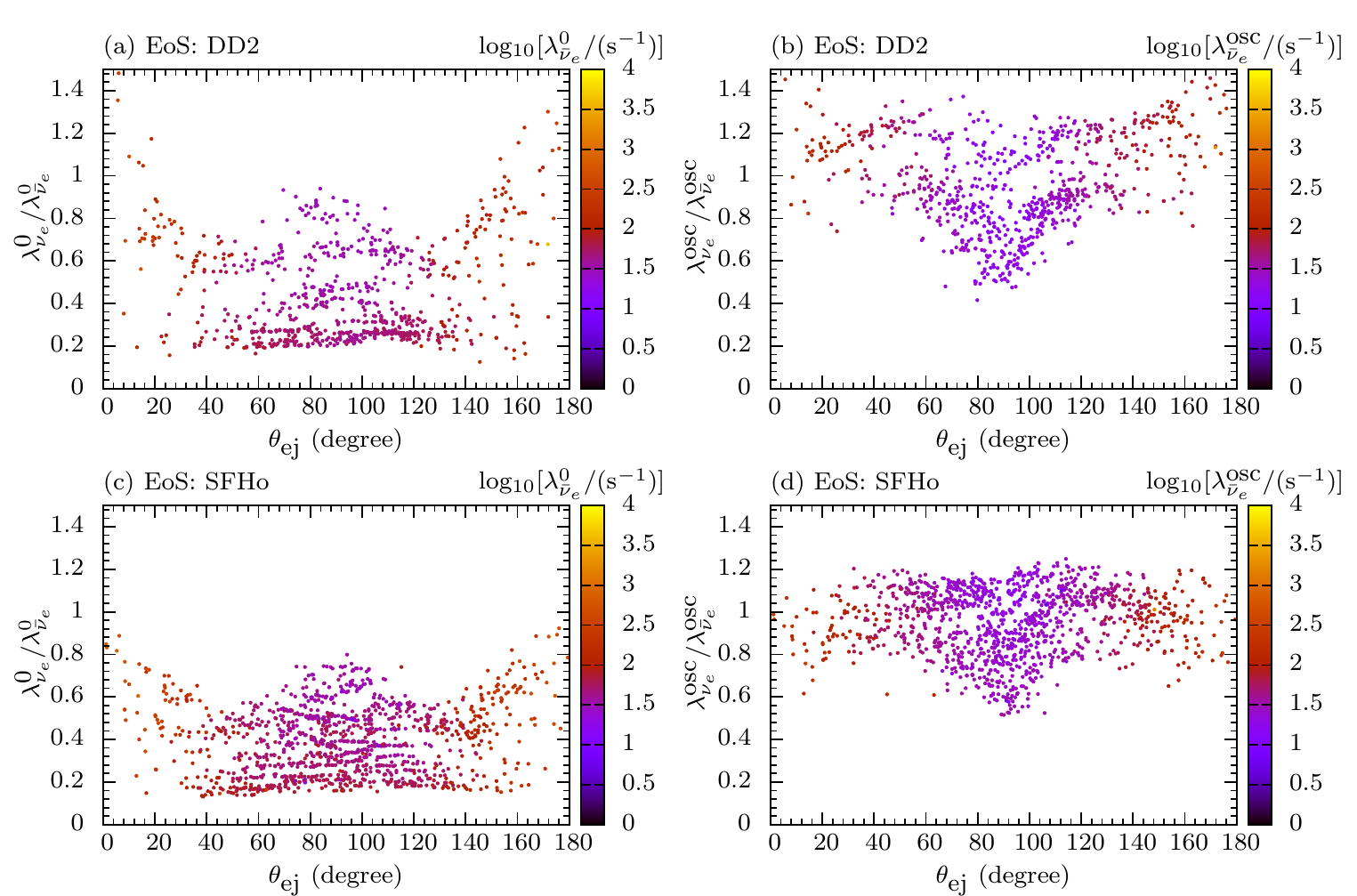}
    \caption{Distribution of the ratios of the $\nu_e$ absorption rates on neutrons, $\lambda_{\nu_e}$, to the $\bar\nu_e$ absorption rates on protons, $\lambda_{\bar\nu_e}$, as a function of $\theta_{\rm ej}$ for all tracer particles at $r=100$~km in the $1.35+1.35$~$M_\odot$ models.
    Panels (a) and (c) are  without flavor conversions, while panels (b) and (d) include flavor equipartition. The color of each point indicates the absolute rate of $\lambda_{\bar\nu_e}$.}
    \label{fig:rateratio}
\end{figure*}

We show in Fig.~\ref{fig:rateratio} the comparison of the ratio of $\lambda_{\nu_e}/\lambda_{\bar\nu_e}$ evaluated at $r=100$~km for all tracer particles for the cases with and without flavor equipartition. 
The corresponding values of $\lambda_{\bar\nu_e}$ are also shown.
These figures highlight that the neutrino absorption rates are orders of magnitudes larger in the polar region than close to the equator. 
For the case without neutrino flavor equipartition, nearly all tracer particles have $\lambda^0_{\nu_e}/\lambda^0_{\bar\nu_e} \lesssim 1$, reflecting the stronger $\bar\nu_e$ flux emitted from its surface. 
With flavor equipartition, the nearly equal contribution of the converted $\nu_x$ and $\bar\nu_x$ significantly changes the ratio $\lambda^{\rm osc}_{\nu_e}/\lambda^{\rm osc}_{\bar\nu_e}$ for both EoS. 
For the case with DD2 EoS, most of the trajectories with $\theta_{\rm ej}\lesssim 60^\circ$ or $\theta_{\rm ej}\gtrsim 120^\circ$ have $\lambda^{\rm osc}_{\nu_e}/\lambda^{\rm osc}_{\bar\nu_e}\gtrsim 1$.
On the other hand, the values of $\lambda^{\rm osc}_{\nu_e}/\lambda^{\rm osc}_{\bar\nu_e}$ scatter around 1 for the model with SFHo EoS.

\begin{figure*}[ht]
    \includegraphics[width = 0.8\linewidth]{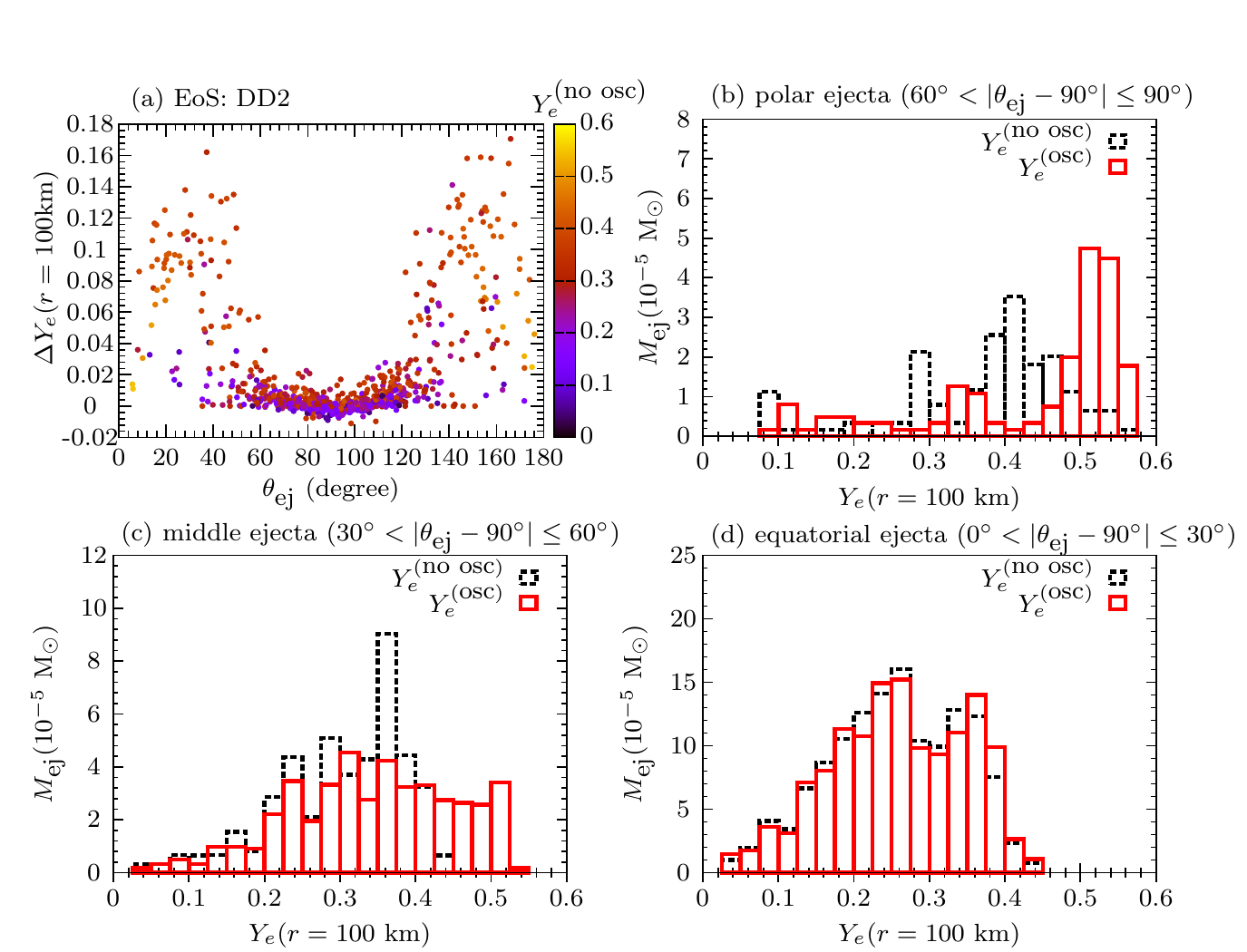}
    \caption{Impact of neutrino flavor equipartition on the ejecta $Y_e$ for the $1.35+1.35$~$M_\odot$ model with DD2 EoS. Panel (a) shows the distribution $\Delta Y_e=Y_e^{\rm (osc)} - Y_e^{\rm (no~osc)}$, at $r=100$~km, as a function of $\theta_{\rm ej}$ for all tracer particles. The color bar represents  $Y_e$ without flavor conversions.
    Panels (b)--(d) show the histograms of the ejecta mass distributions as functions of $Y_e$ at $r=100$~km, for the polar, middle, and equatorial ejecta, respectively.
    }\label{fig:dd2yehisto}
\end{figure*}

\begin{figure*}[ht]
    \includegraphics[width = 0.8\linewidth]{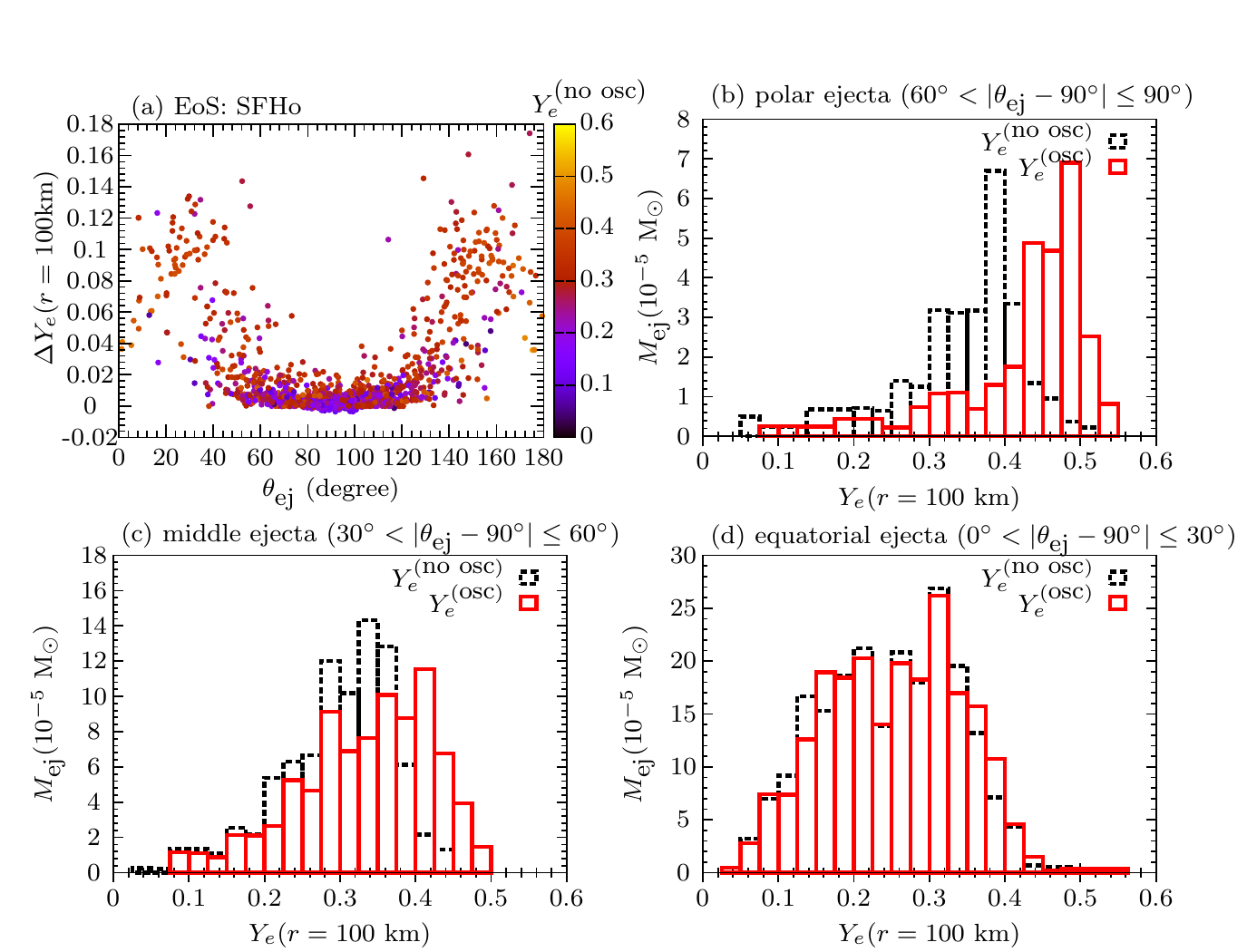}
    \caption{Same as Fig.~\ref{fig:dd2yehisto}, but for the $1.35+1.35$~$M_\odot$ model with SHFo EoS.}\label{fig:sfhoyehisto}
\end{figure*}

The large change in $\lambda_{\nu_e}/\lambda_{\bar\nu_e}$ strongly affects the $Y_e$ distribution of the ejecta.
Figures~\ref{fig:dd2yehisto}(a) and \ref{fig:sfhoyehisto}(a) show  $\Delta Y_e\equiv Y_e^{({\rm osc})}-Y_e^{({\rm no~osc})}$ calculated at $r=100$~km for each tracer particles as a function of the corresponding $\theta_{\rm ej}$, 
where $Y_e^{({\rm osc})}$ and $Y_e^{({\rm no~osc})}$ are the $Y_e$ values with and without flavor equipartition.
These results show that the flavor equipartition can significantly increase  $Y_e$ of the ejecta up to $\sim 0.15$ for  $\theta_{\rm ej}<60^\circ$ or $\theta_{\rm ej}>120^\circ$ closer to the polar directions.
In particular, the increase of $Y_e$ due to flavor equipartition for these ejecta is more pronounced with $Y_e^{\rm (no~osc)}\simeq 0.3-0.4$.
For the ejecta with $Y_e^{\rm (no~osc)}\lesssim 0.2$ or $Y_e^{\rm (no~osc)}\simeq 0.5$, 
 $Y_e$ is less affected.
This is because the ejecta with $Y_e^{\rm (no~osc)}\lesssim 0.2$ expand too fast for neutrino absorption to raise  $Y_e$ either with or without flavor conversion. On the other hand, for the ejecta  with $Y_e^{\rm (no~osc)}\simeq 0.5$,  $\lambda_{\nu_e}/\lambda_{\bar\nu_e} \simeq 1$ even without flavor conversions.
As for the tracer particles with $60^\circ\leq \theta_{\rm ej}\leq 120^\circ$ closer to the disk mid-plane, $Y_e$ is barely influenced by flavor equipartition because of the low neutrino absorption rates (see Fig.~\ref{fig:rateratio}).

Figures~\ref{fig:dd2yehisto}(b)-(d) and \ref{fig:sfhoyehisto}(b)-(d) further show the comparison of the $Y_e$ distribution for the cases with and without flavor conversion, for the ejecta classified into three groups according to  $\theta_{\rm ej}$: the polar ejecta with $60^\circ< |\theta_{\rm ej}-90^\circ|\leq 90^\circ$, the middle ejecta with $30^\circ< |\theta_{\rm ej}-90^\circ|\leq 60^\circ$, and
the equatorial ejecta with $0^\circ\leq |\theta_{\rm ej}-90^\circ|\leq 30^\circ$.
Flavor equipartition influences the $Y_e$ distribution of the polar ejecta by shifting the peak from $\sim 0.4$ to $\sim 0.55$ (0.5) for the DD2 (SFHo) model.
The larger (smaller) shift of the $Y_e$ peak in the DD2 (SFHo) model is related to the larger (smaller) values of $\lambda^{\rm osc}_{\nu_e}/\lambda^{\rm osc}_{\bar\nu_e}$ shown in Fig.~\ref{fig:rateratio}.
For the middle ejecta, a fraction of ejecta originally with $0.3\lesssim Y^{(\rm no~osc)}_e\lesssim 0.4$ has $Y_e^{\rm (osc)} \simeq 0.4-0.5$ when flavor equipartition is reached, while the distribution with $Y_e\lesssim 0.3$ is barely altered.
As for the equatorial ejecta, the corresponding $Y_e$ distribution is only affected negligibly, as expected.

The impact of flavor equipartition in the neutrino-driven ejecta studied in Ref.~\cite{Wu:2017drk} is to lower $Y_e$ (see Fig.~11 therein), while $Y_e$ increases in the models investigated in this work, as discussed above. 
The main difference is that flavor equipartition leads to a larger reduction of $\lambda_{\nu_e}$ and $\lambda_{\bar\nu_e}$ in the BH--torus case, due to the vanishingly small $\nu_x$ fluxes. 
Moreover, a significant part of the ejecta in the BH--torus case is ejected on timescales of several tens of milliseconds during which the neutrino luminosities decrease substantially (see Figs.~2 and 8 in Ref.~\cite{Wu:2017drk});  this leads to much smaller neutrino absorption rates even without assuming neutrino flavor equipartition. As a consequence, since the ejecta start out as neutron-rich material, a largely reduced $Y_e$ for the neutrino-driven outflow was found  in the BH--torus model of Ref.~\cite{Wu:2017drk}.

\begin{figure}[ht]
    \includegraphics[width = \linewidth]{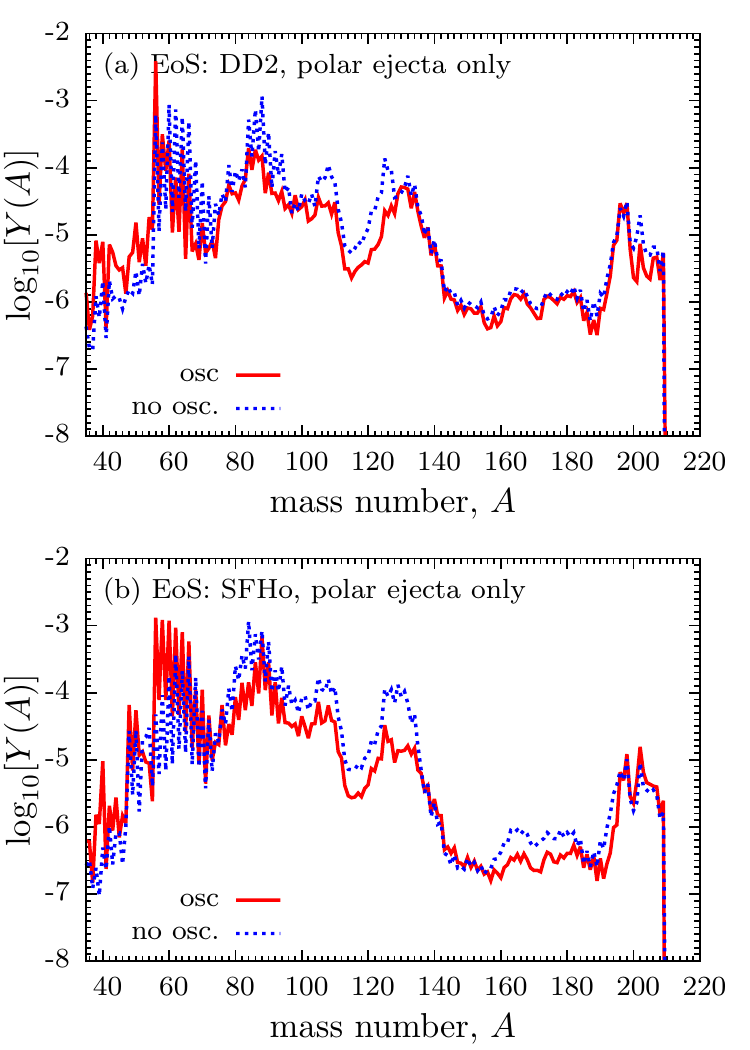}
    \caption{Nucleosynthesis outcome in the polar ejecta for the cases with and without flavor equipartition for the $1.35+1.35$~$M_\odot$ model with  DD2 EoS (a) and for the SHFo EoS model (b).}
    \label{fig:nucleo}
\end{figure}

We show in Fig.~\ref{fig:nucleo} the impact of flavor equipartition on the abundance distribution for the polar ejecta ($60^\circ\leq |\theta_{\rm ej}-90^\circ|\leq 90^\circ$) for the $1.35+1.35$~$M_\odot$ model with both  EoSs. 
Since  flavor equipartition mainly shifts the distribution of high $Y_e\gtrsim 0.3$ material,  noticeable changes appear  regarding the iron peak and the first peak elements. In particular, the amount of produced $A=56$ nuclei is enhanced by a factor of $\sim 6(27)$ for the DD2(SFHo) EoS model. 
The amount of lanthanides in the polar ejecta, relevant to the kilonova color, is affected negligibly for the  DD2 model and reduced by $\sim$ a factor of 2 for the SHFo model. 
Since the polar ejecta contribute up to $\sim 10\%$ to the total ejecta mass considered here, the overall modifications induced by flavor equipartition on the total abundance yields are relatively small. 
For instance, the change of the produced amount of strontium at the time of a day and the amount of free neutrons at the end of the $r$-process is at the level of $\sim 10\%$.

\section {Summary and discussion}
\label{sec:conclusion}

In this work, we have examined the neutrino emission properties and the conditions for the occurrence  of 
fast neutrino flavor conversions during the first 10~ms after
the coalescence of symmetric ($1.35+1.35$~$M_\odot$) and asymmetric ($1.25+1.45$~$M_\odot$) NS binaries, during which the remnant consists of a hypermassive NS 
surrounded by an accretion disk. 
We have also performed detailed analyses regarding the properties of the material ejected during this phase and nucleosynthesis calculations for cases with and without neutrino flavor mixing.
Our study is based on the outputs from the  
general-relativistic simulations with approximate neutrino transport, performed with two different EoS (DD2 and SHFo) based on Refs.~\cite{Ardevol:thesis,Ardevol-Pulpillo:2018btx} and available at~\cite{Ardevol2019}.

Flavor instabilities that may lead to fast pairwise flavor conversions occur throughout
the whole investigated post-merger evolution, independent of the adopted EoS or the mass ratio of the binary.
This is a direct consequence of the $\bar\nu_e$ emission dominating over the $\nu_e$ one due to the protonization of the merger remnant,
which leads to crossings in the neutrino ELN angular distributions
everywhere above the neutrino emitting surfaces.
Our results thus confirm the earlier conclusions of Ref.~\cite{Wu:2017qpc}, which adopted
a simple toy-model for the  neutrino emission characteristics and emission surface geometry.
However, in contrast to the results obtained in Ref.~\cite{Wu:2017drk}, which showed that the
region where flavor instabilities exist shrinks on a time scale of $\sim \mathcal{O}(10)$~ms as the BH--disk remnant evolves, 
the flavor unstable regions reported here remain quite stable within the examined $10$~ms of post-merger evolution.
Since Refs.~\cite{Perego:2014fma, Fujibayashi:2017puw}
reported dominating emission of $\bar\nu_e$ over $\nu_e$ on time scales 
longer than $\mathcal{O}(100)$~ms
for a hypermassive NS accretion disk system, 
we expect that the flavor instabilities found in this paper may be sustained for even longer duration 
and affect the nucleosynthesis in the disk winds.

As for the ejecta properties and the nucleosynthesis outcome, the ejecta contain a wide $Y_e$ distribution up to 0.5 due to the effect of weak interactions including neutrino absorption, allowing for the formation of heavy elements in all three $r$-process peaks. 
In particular, a few times $10^{-5}$~$M_\odot$ of strontium are synthesized in the ejecta in all models, consistent with the amount inferred from the GW170817 kilonova observation~\cite{Watson:2019xjv}. 
We also find that the amount of free neutrons  left after the  $r$-process freeze-out  is roughly a factor of $10$ smaller than the one obtained in simulations without taking into account the effect of weak interactions. This has implications for the prediction of the early-time UV emission that may be powered by the decay of free neutrons~\cite{Metzger:2014ila}.

By relying on the extreme ansatz that fast pairwise conversions lead to  flavor equilibration, we find that flavor mixing of neutrinos mostly affects the polar ejecta within $\sim 30^\circ$ by changing the peak $Y_e$ from $\sim 0.4$ to $\sim 0.5$. The dominant effect is thus to reduce the first peak abundances while enhancing the iron peak abundances.
This result quantifies the most extreme impact of neutrino flavor conversions on the nucleosynthesis in the early-time ejecta; most likely, the effect due to neutrino flavor conversions should be between our explored two cases with and without oscillations. 
Note that we have only examined the flavor instability above the neutrino emitting surfaces. Beyond that, future work investigating the occurrence of unstable conditions inside the neutrino-trapping regime,
along the lines of recent work done in the context of core-collapse supernovae~\cite{Abbar:2019zoq,DelfanAzari:2019tez,Morinaga:2019wsv,Nagakura:2019sig,Glas:2019ijo,Johns:2019izj,Abbar:2020fcl}, should be carried out.
Potential effects due to the presence of turbulence~\cite{Abbar:2020ror} may also play a role in post-merger environments.

Multidimensional numerical simulations tracking the flavor evolution in the presence of  fast pairwise conversions  (see, e.g., 
Refs.~\cite{Abbar:2018beu,Martin:2019gxb,Shalgar:2019qwg,Bhattacharyya:2020dhu,Capozzi:2020kge,Shalgar:2020xns,1815221}), including the  collisional term in the equations of motion, are essential to draw
robust conclusions on the role of neutrino flavor conversion 
for the outcome of nucleosynthesis in the ejecta and the corresponding kilonova observables.

\begin{acknowledgments}
MG and MRW acknowledge support from the Ministry of Science and Technology, Taiwan under Grant No.~107-2119-M-001-038, No.~108-2112-M-001-010, No.~109-2112-M-001-004, 
the Academia Sinica under project number AS-CDA-109-M11, 
and the Physics Division of National Center for Theoretical Sciences.
IT acknowledges support from the Villum Foundation (Project No.~13164), the Danmarks Frie Forskningsfonds (Project No.~8049-00038B), the Knud H\o jgaard Foundation.
At Garching, funding by the European Research Council through grant ERC-AdG No. 341157-COCO2CASA and by
the Deutsche Forschungsgemeinschaft (DFG, German Research Foundation) through grants SFB-1258 ``Neutrinos and Dark Matter in Astro- and Particle Physics (NDM)'' and under Germany's Excellence Strategy through Excellence Cluster ORIGINS (EXC 2094)---390783311 is acknowledged.

\end{acknowledgments}

\appendix
\section{Computing the neutrino number densities from the simulation data}\label{app:postproc}

As the simulation outputs from Refs.~\cite{Ardevol-Pulpillo:2018btx,Ardevol:thesis} did not directly store the $\nu_e$ and $\bar\nu_e$ absorption rates on free nucleons along the ejecta trajectories, we compute the rates in a post-processing fashion as follows. 
First, the simulations provide the local $\nu_e$ and $\bar\nu_e$ energy luminosities and average energies at radii of 50 and 100~km as functions of the polar angle $\theta$ (with respect to the $z$-axis) for different  post-merger time snapshots.
This allows to compute the $\nu_e$ and $\bar\nu_e$ number densities at these two radii.
Then, at a given time $t$, for any spatial coordinate $\bm{x}$ on a trajectory with  radius 50~km$\leq r\leq 100$~km (with an angle $\theta$), we compute the corresponding number densities of $\nu_e$ and $\bar\nu_e$ by linearly interpolating the logarithmic values of the densities obtained above at $r=50$ and $100$~km. 
Once we have the number densities of the $\nu_e$ and $\bar\nu_e$, the absorption rates for $50$~km$\leq r\leq 100$~km can  be computed using Eqs.~\eqref{eq:lamb_nue} and \eqref{eq:lamb_anue} in Appendix.~\ref{app:rates}.
For $r>100$~km, we extrapolate the rates assuming that they scale as $r^{-2}$:
\begin{equation}
    \lambda_{\nu_\alpha}(r>100{\rm~km})=
    \lambda_{\nu_\alpha}(r=100{\rm~km})\times\left(\frac{\rm 100~km}{r}\right)^2,
\end{equation}
For $r<50$~km, we take a different form of extrapolation to partly account for the finite-size emission geometry:
\begin{eqnarray}\label{eq:innerextra}
    \lambda_{\nu_\alpha}(r<50{\rm km})&=&
    \lambda_{\nu_\alpha}(r=50{\rm km})\times\\ \nonumber
   & &\left(\frac{1-\sqrt{1-(r_0/r)^2}}{1-\sqrt{1-(r_0/(50{\rm km}))^2}}\right)^2,
\end{eqnarray}
with $r_0=25$~km.
We have checked that even simply taking the $1/r^2$ extrapolation for regions with $r<50$~km leads to nearly identical results to those obtained by using Eq.~\eqref{eq:innerextra}.

\begin{figure}[t]
    \includegraphics[width = \linewidth]{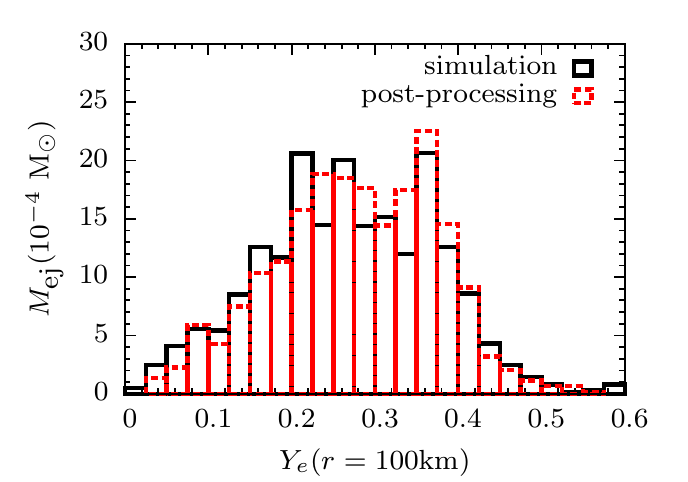}
    \caption{Ejecta $Y_e$ histograms computed via post-processing and as from the  simulations of Refs.~\cite{Ardevol-Pulpillo:2018btx,Ardevol:thesis}.}
    \label{fig:yecomp}
\end{figure}

Figure~\ref{fig:yecomp} compares the $Y_e$ distribution at $r=100$~km, obtained by the simulation of Ref.~\cite{Ardevol-Pulpillo:2018btx} to the one obtained by using the above neutrino absorption rates in the nuclear reaction network described in Sec.~\ref{sec:nuabsorb}.
It shows that the $Y_e$ distributions agree with each other reasonably well. 

For $\nu_x$, since the simulations do not store the energy luminosity and average energy in the bins at 50 and 100~km, we  estimate the $\nu_x$ number density and average energy along each ejecta trajectory as follows. 
First, the locations of the $\nu_x$ surface and the associated temperatures for times between $2.5\leq t\leq 10$~ms are interpolated by using the data provided at $2.5, 5, 7.5$ and $10$ ms.
Then, for a given time $t$, a $\nu_{x}$ number density on each point $\vec x$, ${\tilde n}_{\nu_x}(\vec x)$, at the emission surface following the Fermi-Dirac distribution with temperature $T(\vec x_i)$ and zero chemical potential can be easily calculated. 
We then re-normalize the total neutrino number luminosity emitted from the $\nu_x$ surface to the value given by simulation data,
\begin{equation}
    L_{N,\nu_x}=\frac{L_{\nu_x}}{\langle E_{\nu_x}\rangle} = \frac{5\cdot \xi}{12} \int dS {\tilde n}_{\nu_x},
\end{equation}
where $L_{\nu_x}$ and $\langle E_{\nu_x}\rangle$ are the energy luminosity and mean energy of $\nu_x$ shown in Fig.~\ref{fig:lumegy}, respectively. 
The quantity $dS$ is the differential surface area on the $\nu_x$ surface. The factor $5/12$ accounts for the forward peaked angular profile of $\nu_x$ emission consistent with Eq.~\eqref{emission_spectrum}, and $\xi$ is the normalization constant.
Correspondingly, the rescaled $\nu_x$ number density on their emission surface is given by $n_{\nu_x}(\vec x)=\xi{\tilde n}_{\nu_x}(\vec x)$.

We assume the $\nu_x$ average energy at each location $\vec x$ on the emission surface to be 
\begin{equation}
    \langle E_{\nu_x}\rangle(\vec x) = \frac{1}{2}\left(\langle E_{\nu_x}\rangle + 3.15~T(\vec x)\right),
\end{equation}
to partly account for the fact that the $\nu_x$-$e^\pm$ scatterings, which can down-scatter $\nu_x$, was not included in the numerical simulations of Ref.~\cite{Ardevol-Pulpillo:2018btx}.
Since the local temperature on the $\nu_x$ emission surface within $x\lesssim 20$~km is found to be $\gtrsim 6$~MeV, the main effect of the above choice is to reduce the $\langle E_{\nu_x}\rangle(\vec x)$ at the outer edge of the emission surface. We have additionally confirmed that adopting a location independent average energy of $\nu_x$  given by $\langle E_{\nu_x}\rangle$ 
does not qualitatively change our results shown in the main text.

For  $t<2.5$~ms, we simply assume that the neutrino emission surface is the same as the one at $t=2.5$~ms, and scale the number density $n_{\nu_x}(\vec x,t)$ and the average energy $\langle E_{\nu_x}\rangle(\vec x,t)$ in the following way
\begin{align}
    n_{\nu_x}(\vec x,t) &= n_{\nu_x}(\vec x,t=2.5~{\rm ms})\left( \frac{L_{N,\nu_x}(t) }{L_{N,\nu_x}(t=2.5~{\rm ms})}\right),\\
    \langle{E_{\nu_x}}\rangle(\vec x,t) &= \langle{E_{\nu_x}}\rangle(\vec x,t=2.5~{\rm ms})\left( \frac{\langle E_{\nu_x}\rangle(t) }{\langle E_{\nu_x}\rangle(t=2.5~{\rm ms})} \right).
\end{align}

Once we have the desired quantities on the emission surface for all times, we  use the same ray-tracing technique as in the main text to compute the $\nu_x$ number densities for the locations crossed by the trajectories. 
The absorption rates of the converted $\nu_x$ and $\bar\nu_x$ on nucleons, $\lambda_{\nu_x}$ and $\lambda_{\bar\nu_x}$, along all trajectories, are similarly computed as those of $\nu_e$ and $\bar\nu_e$ given in Appendix~\ref{app:rates} by replacing the corresponding number densities, the average energies, and other higher energy moments.

\section{Computing the neutrino absorption rates}\label{app:rates}
In order to compute the evolution of $Y_e$ for the outflows for the cases with and without flavor conversions, we first compute the number densities and average energies of $\nu_e$, $\bar\nu_e$ and $\nu_x$ (without flavor conversions) along the trajectories of all tracer particles by post-processing the simulation data as detailed in Appendix~\ref{app:postproc}.
For the case without flavor conversions, we  follow Ref.~\cite{Pllumbi:2014saa} to calculate the $\nu_e$ and $\bar\nu_e$ absorption on free nucleons:
\begin{align}
    \lambda_{\nu_e}^0 &= n_{\nu_e} \langle\sigma_{\nu_e}\rangle,\label{eq:lamb_nue}\\
    \lambda_{\bar\nu_e}^0 &= n_{\bar\nu_e} \langle\sigma_{\bar\nu_e}\rangle,
    \label{eq:lamb_anue}
\end{align}
where $\langle\sigma_{\nu_e}\rangle$ and $\langle\sigma_{\nu_e}\rangle$ are the spectrally averaged absorption cross-sections of $\nu_e$ and $\bar\nu_e$. 
By taking into account the recoil corrections and weak magnetism~\cite{Horowitz:1999fe}, the average neutrino capture cross sections are approximated by
\begin{align}
    \langle\sigma_{\nu_e}\rangle &\simeq k\langle E_{\nu_e} \rangle \varepsilon_{\nu_e} \left[1+2\left(\frac{\Delta}{\varepsilon_{\nu_e}}\right)+a_{\nu_e}\left(\frac{\Delta}{\varepsilon_{\nu_e}}\right)^2\right]W_{\nu_e},\label{eq:sigma_nu}\\
    \langle\sigma_{\bar\nu_e}\rangle &\simeq k\langle E_{\bar\nu_e} \rangle \varepsilon_{\bar\nu_e} \left[1+2\left(\frac{\Delta}{\varepsilon_{\bar\nu_e}}\right)+a_{\bar\nu_e}\left(\frac{\Delta}{\varepsilon_{\bar\nu_e}}\right)^2\right]W_{\bar\nu_e},\label{eq:sigma_anu}
\end{align}
where $k=9.3\times10^{-44}$~cm$^2/$MeV$^2$, $\varepsilon_{\nu_e,\bar\nu_e}=\langle E^2_{\nu_e,\bar\nu_e} \rangle/\langle E_{\nu_e,\bar\nu_e}\rangle$, $a_{\nu_e,\bar\nu_e} = \langle E^2_{\nu_e,\bar\nu_e}\rangle/\langle E_{\nu_e,\bar\nu_e}\rangle^2$, 
and $\Delta=(m_n-m_p)=1.293~\tx{MeV}$ is the neutron-proton mass difference. 
The weak-magnetism and recoil correction factors $W_{\nu_e,\bar\nu_e}$ are given by
\begin{eqnarray}
    W_{\nu_e} &=& \left[1+1.02\frac{b_{\nu_e}\varepsilon_{\nu_e}}{M}\right],\\
    W_{\bar\nu_e} &=& \left[1-7.22\frac{b_{\bar\bar\nu_e}\varepsilon_{\bar\nu_e}}{M}\right],
\end{eqnarray}
where $b_{\nu_e,\bar\nu_e}=\langle E^3_{\nu_e,\bar\nu_e} \rangle\langle E_{\nu_e,\bar\nu_e}\rangle/\langle E^2_{\nu_e,\bar\nu_e} \rangle^2$
is the spectral shape factor for $\nu_e(\bar\nu_e)$
and $M=940$ is roughly the mass of a nucleon in MeV.
Note that in deriving the rates through the above equations, we have assumed zero chemical potentials for all neutrino species to compute the $i$-th neutrino energy moments $\langle E_{\nu_\alpha}^i\rangle$.

%

\end{document}